\documentclass{article}
\usepackage{setspace}
\usepackage{graphicx} 
\usepackage{hyperref}
\hypersetup{
    colorlinks,
    linkcolor={blue},
    citecolor={blue},
    urlcolor={blue}
}
\usepackage{amsmath,amssymb,amsthm}
\usepackage{lineno}
\usepackage[capitalise]{cleveref}
\usepackage{subcaption}

\newtheorem{theorem}{Theorem}
\newtheorem{lemma}[theorem]{Lemma}

\newtheorem{corollary}{Corollary}[theorem]
\newtheorem*{remark}{Remark}

\usepackage{amsfonts}
\usepackage{xcolor}
\usepackage{enumitem}
\usepackage{tkz-graph}  
\usetikzlibrary{shapes.geometric}
\usetikzlibrary{positioning,fit}

\usepackage{xcolor}
\usepackage{caption}
\usepackage{tabularray}
\UseTblrLibrary{booktabs}
\usepackage{xspace}

\usepackage[authoryear, round, compress]{natbib}

\newcommand{\R}{\mathbb{R}}
\newcommand{\E}{\mathbb{E}}
\newcommand{\N}{\mathcal{N}}
\newcommand{\pr}{\mathbb{P}}

\newcommand{\I}[1]{\mathbb{I}[#1]}
\newcommand{\Id}{\mathbf{I}}

\newcommand{\Hs}{\mathcal{H}}

\newcommand{\vv}{\mathcal{V}}
\newcommand{\vvk}{\vv_k}
\newcommand{\vvmk}{\vv_{(-k)}}
\newcommand{\betaestk}{\hat{\beta}_k(w)}

\newcommand{\estPik}{p_{a}(X_i; \betaestk)}
\newcommand{\estPiki}{p_{a}(X_i; \betaestk)}
\newcommand{\truePi}{p_a(X_i; \beta_w)}
\newcommand{\ytild}{\tilde{Y}_i}
\newcommand{\ytilds}{\tilde{Y}_i^w}
\newcommand{\derivPiki}{\frac{\partial}{\partial \beta} p_w(X_i; \beta) \bigg |_{\beta=\beta_w}}

\newcommand{\estHiok}{\hat{h}_{1,k}(X_i)} 
\newcommand{\trueHio}{\tilde h_{1,k}(X_i)}
\newcommand{\estHizk}{\hat{h}_{0,k}(X_i)} 

\newcommand{\trueMo}{\mu_1}
\newcommand{\trueMz}{\mu_0}
\newcommand{\maSi}[1]{m_{#1}(W_i)}
\newcommand{\gaSi}[1]{g_{#1}(W_i)}
\newcommand{\mas}[1]{m_{#1}(w)}
\newcommand{\gas}[1]{g_{#1}(w)}
\newcommand{\Sg}{\mathcal{W}}

\newcommand{\hesttmlei}[1]{\hat{h}^{(u)}_{#1,k}(X_i; \hat{\epsilon}, \betaestk)}
\newcommand{\hesttmlestari}[1]{\hat{h}^{(u)}_{#1,k}(X_i; \tilde \epsilon, \beta_w)}
\newcommand{\htruetmlei}[1]{\tilde h^{(u)}_{#1,k}(X_i; \tilde \epsilon, \beta_w)}
\newcommand{\hbaseest}[1]{\hat{h}^{(u)}_{#1,k}}

\newcommand{\opo}{o_P(1)}
\newcommand{\Opo}{O_P(1)}

\newcommand{\prob}{\mathbb{P}}
\newcommand{\V}{\mathbb{V}}

\newcommand{\isS}{\I{W_i=w}}

\newcommand{\Sn}{W^{(n)}}
\newcommand{\An}{A^{(n)}}

\newcommand{\Vmisspec}{\Sigma_{misspec}}

\newcommand{\qt}{\quad\quad\quad}

\newcommand{\RD}{\texttt{RD}}
\newcommand{\RMST}{\texttt{RMST}}
\newcommand{\RR}{\texttt{RR}}
\newcommand{\WR}{\texttt{WR}}
\newcommand{\OR}{\texttt{OR}}

\newcommand{\Surv}[2]{S(#1, #2)}
\newcommand{\St}[1]{\Surv{t}{#1}}

\newcommand{\Sta}{\Surv{t}{a}}

\newcommand{\Survest}[2]{\hat{S}(#1, #2)}

\newcommand{\Saest}[1]{\hat{S}(#1, a, X)}
\newcommand{\Staest}{\Survest{t}{a}}

\newcommand{\Vta}{V_{t,a}}

\newcommand{\IF}{\phi}
\newcommand{\IFta}{\phi_{t,a}}

\newcommand{\pa}{p_A}
\newcommand{\pc}{p_C}
\newcommand{\paest}{\hat{p}_A}
\newcommand{\pcest}{\hat{p}_C}

\newcommand{\hesttax}{\hat{h}(t, a, X)}

\newcommand{\hesttaxtmle}{\hat{h}^{\mbox{tmle}}(t, a, X)}

\newcommand{\Ptrue}{\mathsf{P}}

\newcommand{\Gn}{\mathsf{G}_n}

\newcommand{\Lpw}{L^2(P^w)}

\newcommand{\rck}[1]{\textcolor{blue}{RCK: #1}}
\usepackage[most]{tcolorbox}
\usepackage{xcolor}

\definecolor{nyupurple}{HTML}{57068C}

\usepackage{authblk}
\providecommand{\tightlist}{%
\setlength{\itemsep}{0pt}\setlength{\parskip}{0pt}}

\newcommand{\anon}{0}

\newcommand{\titleString}{Data-Adaptive and Model-Robust Covariate Adjustment for Time-to-Event Outcomes in Stratified Trials}

\providecommand{\tightlist}{%
  \setlength{\itemsep}{0pt}\setlength{\parskip}{0pt}}\usepackage{longtable,booktabs,array}
\usepackage{calc} 

\begin{document}
\ifnum\anon=0
{
  \title{\Large \titleString}
  \date{}
  \author[1,2]{Raphael C. Kim}
  \author[2]{Brian Gilbert}
  \author[1]{Ramin Zabih}
  \author[2]{Michele Santacatterina}
  \author[2]{Ivan Diaz}
  \affil[1]{Cornell Tech, Cornell University (New York, NY)}
  \affil[2]{Department of Biostatistics, Department of Population Health, New York University School of Medicine (New York, NY)}
  \maketitle
} \fi

\ifnum\anon=1
{
  \bigskip
  \bigskip
  \bigskip
  \begin{center}
      {\LARGE\bf \titleString}
\end{center}
  \medskip
} \fi

\begin{abstract}
    Time-to-event outcomes are commonly used as primary endpoints in randomized clinical trials. Despite this, relatively little work incorporates baseline covariate information while also accounting for stratified randomization, a common form of randomization. Moreover, leveraging efficiency gains using these approaches typically requires pre-specifying a subset of covariates that are most predictive of the outcome—a challenging task in practice, as most trials collect dozens of potentially prognostic baseline variables. In this work, we build on existing literature to propose a data-adaptive and model-robust covariate adjustment method for time-to-event outcomes. Our approach, based on targeted minimum loss-based estimation, allows for data-adaptive covariate selection and model-robust efficient inference on functionals of the survival curve while accounting for stratification. Through extensive simulations and analysis, we showcase the simplicity and improved precision of our method when the covariate set is not known \textit{a priori}.
\end{abstract}

\small{\textit{Keywords: } Covariate Adjustment, Time-to-event, Clinical Trials, Targeted Minimum Loss-based Estimation, Data-Adaptive Estimation}


\newpage 

\section{Introduction}

Time-to-event analysis serves as a primary objective in many randomized clinical trials (RCTs). 
To improve statistical precision, RCTs routinely employ two strategies: stratified randomization, in which treatment is assigned within levels of key baseline covariate levels, and covariate adjustment, which accounts for chance imbalances in treatment groups using baseline covariates \citep{Tsiatis08}. 

Despite the ubiquity of time-to-event analyses, methods that optimally incorporate stratification and covariate adjustment are limited. Under a proportional hazards assumption, one may employ a stratified Cox proportional hazards model \citep{KleinbaumKlein12}, which allows the baseline hazard to vary across strata while estimating a common treatment effect. Although this approach accounts for stratification and permits covariate adjustment, it has important limitations. First, it targets a hazard ratio, a conditional estimand that may be difficult to interpret clinically \citep{Stensrud19,FayLi24}. Second, it relies on modeling assumptions, including proportional hazards and correct specification of the functional form. When there are several baseline covariates, as is the case in many RCTs, restrictive modeling assumptions are likely to result in misspecification and possible precision loss.
 
Recent advances in causal inference and machine learning have enabled flexible, data-adaptive approaches to covariate adjustment for a wide range of estimands \citep{Williams22, VLDV24}, including various time-to-event estimands involving the survival curve, restricted mean survival time (\RMST), and more. These methods can leverage rich covariate information and reduce reliance on parametric assumptions, but they assume simple randomization, failing to account for stratified designs.
Only recently have modern methods for covariate adjustment under stratified randomization received closer attention. 

\cite{Bugni18} presented valid inferential procedures for analyzing estimands under stratified and biased coin randomization settings. They found that improved precision gains under stratification can be leveraged, although they only allowed the practitioner to adjust for covariates used for stratification. \cite{Wang19} generalized this work to M-estimators, permitting the inclusion of a wider set of covariates for various outcome types and developing unadjusted time-to-event analysis methods under stratification via Kaplan-Meier methods. Similar to the stratified Cox proportional hazards model \citep{KleinbaumKlein12}, the covariate adjustment methods of \cite{Bugni18} and \cite{Wang19} rely on optimally pre-selecting a predictive set of covariates for use in the outcome model to maximize the precision gains. \cite{Rafi23}, \cite{Tu24}, and \cite{Bannick25} address the problem of covariate adjustment under stratified randomization with data-driven covariate selection and flexible, nonparametric estimators, but not for time-to-event outcomes.

Despite the remarkable progress in covariate adjustment and time-to-event analysis, to the best of our knowledge, existing covariate adjustment methods do not have the ability to target general time-to-event estimands, utilize data-driven model building, and leverage stratified randomization designs. Given the widespread use of time-to-event endpoints, this presents a significant methodological gap in the analysis of RCTs.
\\\\
{\textbf{Contributions and Organization:} }

In this work, we propose a Targeted Minimum Loss-based Estimators (TMLE) for data-adaptive and model-robust covariate adjustment on time-to-event outcomes under stratified randomization. Our approach allows the practitioner to target various time-to-event estimands of interest such as the survival curve, \RMST, and more.
In our analyses, we demonstrate precision gains from two sources: (i) data-adaptive covariate selection and modeling flexibility, and (ii) stratified randomization. We also extend TMLE methods to outcome types beyond time-to-event outcomes in \cref{sec:tmle_generalizations}, allowing standard TMLE approaches used under simple randomization to be applied in stratified settings with a simple, variance correction, supporting the usability of our methods.

The rest of our paper is organized as follows.
In \cref{sec:methods}, we present theoretical results showing precision gains under stratification, with an extensive simulation study found in  \cref{sec:simulations} to verify our theoretical properties. In  \cref{sec:rwd}, we illustrate our precision gains through a re-analysis of a rare disease clinical trial in schizophrenia known as the Depression and Citalopram in First Episode Recovery (DECIFER) trial \citep{Goff19}. Finally, we conclude with a discussion in \cref{sec:conclusion}.

\section{Methods}\label{sec:methods}
\textbf{Setup: } We work in a discrete timepoint setting in which we have $t \in [T]$ times. Let $Z_i=(X_i, Y_i(0), Y_i(1), C_i(0), C_i(1)) \overset{i.i.d.}{\sim} P$ be the full data vector with potential outcomes, for baseline covariates $X_i$, potential outcomes $Y_i(0), Y_i(1)$, and potential right-censoring times $C_i(0), C_i(1)$. 

We work in a stratified randomization setting in which treatment $A_i$ conditional on strata $W$. Denote the distribution conditional on the strata by $P^w$, or $P(\cdot \mid W=w)$. As in \cite{Wang19}, we will let the stratification variable be denoted by $W_i$ with $W_i \subset X_i$. Under this setup, we let $O_i = (X_i, A_i, \delta_i, U_i)$ be our observed data tuples, for censoring indictor $\delta_i = \I{Y_i \leq C_i}$, and observed outcome $U_i = \min(Y_i, C_i)$.  Further, let $\An = (A_1 \dots A_n)$, $\Sn=(W_1 \dots W_n)$, $O^{(n)} = (O_1 \dots O_n)$, and $Z^{(n)}=(Z_1 \dots Z_n)$, the collection of all $n$ data samples of the respective variable. We assume the following \textbf{Identification Assumptions.}
\begin{enumerate}[label=\bfseries (Id\arabic*)]    
    \item \textbf{Consistency.} $Y_i = Y(A_i) = A_i Y_i(1) + (1-A_i) Y_i(0)$ and $C_i = C(A_i) = A_i C_i(1) + (1-A_i) C_i(0)$. \label{ass:consistency}
    \item \textbf{Positivity among strata.} $\prob(A_i =a \mid W_i=w) = p_a \geq \rho, \forall w, a$, where $\rho \in (0, 1)$.
    \label{ass:pos}
    \item \textbf{Stratified Randomization:}  \label{ass:sampling}
        Suppose we have a common stratified randomization probability $p \equiv p_a$ for all $w$. More formally,
        \begin{align*}
        \text{(a)}\quad &\text{Stratified Randomization: } O^{(n)} \perp\!\!\!\perp A^{(n)} \mid \Sn \\[6pt]
        \text{(b)}\quad &\text{Strong Balancing} \\
        &  \forall w \in \Sg, \frac{1}{\sqrt{n}} \sum_{i=1}^n (A_i - p)\,\mathbf{1}\{W_i=w\} = \opo
        \end{align*}
    \item \textbf{Censoring completely at random:}  \label{ass:censoring}
        For each treatment arm $a$, the censoring time is independent of both the outcome and baseline covariates: $C \perp (Y, X) \mid A = a $
\end{enumerate}
It will be helpful to define additional quantities we wish to estimate: 
\begin{enumerate}[topsep=0pt, itemsep=0pt]
    \item Outcome model $h$, or the conditional hazard
    $$
        h(t,a,X_i)  = \prob(U_i=t, \delta_i=1 \mid U_i \geq t, A_i=a, X_i)
    $$
    \item Treatment propensity score model. $\pa(X_i)  = \prob(A_i=1 \mid X_i)$
    \item Censoring propensity score model. $\pc(t, a, X_i)  = \prob(U_i = t, \delta_i = 0 \mid U_i \geq t, A_i=a, X_i)$
    \item Conditional Survival Function. Under our identification assumptions, namely \labelcref{ass:sampling}-\labelcref{ass:censoring}, we have
    $S(t,a,X) = \pr[Y > t \mid A=a, X=x] = \prod_{t=1}^{T} \{1 - h(t,a,X)\} $
    \item Conditional Censoring Function. Similarly, $ \Pi_C(t,a,X) = \pr[C \geq t \mid A=a, X=x] = \prod_{t=1}^{T} \{1 - p_c(t,a,X)\} $
\end{enumerate}

We begin in \cref{sec:survival_fx} by presenting results targeting the survival function at timepoint $t$ under treatment $a$, $\Sta = 1-F(t,a)$ where $F(t,a)=\int \prob[Y \leq t \mid A=a, W=w] d\pr(w)$. With this in hand, we may proceed to consider smooth functionals of $\Sta$ in \cref{sec:functionals} such as the Restricted Mean Survival Time, or Win Ratio; we introduce examples closer in \cref{tab:functional_estimands}.

\subsection{Survival Function}\label{sec:survival_fx}
Our TMLE procedure for targeting $\Sta$ under stratified randomization works as follows. This estimation steps closely follow that of \citet{Williams22}.

\begin{tcolorbox}[width=1.1\linewidth,float=ht!,title=TMLE procedure for the Survival Curve]\label{proc:surv}
\begin{enumerate}
    \tightlist
    \item Construct an estimate for models:
    \begin{itemize}
        \item Outcome model $\hesttax$. This can include data-adaptive variable selection methods.
        \item Treatment propensity score model $\paest(a,X_i) = \pr(A_i=1 \mid X_i; \hat \beta)$ among each stratum variable in $\Sg$, a parametric model in $\hat \beta$
        \item Censoring propensity score model $\pcest(t, a, X_i) = \hat{\pr}(U_i=t, \delta_i=0 \mid U_i \geq t, A_i=a, X_i)$
    \end{itemize}
    \item Compute the clever covariate
    \begin{equation}
        H_t = -\frac{\I{A=a}}{\paest(a, X)\hat{\Pi}_C(t, a, X)} \frac{\Saest{T}}{\Saest{t}}
    \end{equation}
    \item Update the initial estimates by solving for $\varepsilon$ in the model below
    \begin{align*}
        \pr[U_i = t, \delta_i=1 \mid U_i \geq t, A_i=a, X_i] = \mbox{logit}^{-1}(\mbox{logit}(\hesttax)+ \varepsilon H_t)
    \end{align*}
    This yields the estimate $\hesttaxtmle$
    \item Construct an estimator of $\Sta$ by  
    \begin{equation}\label{eq:estSurv}
        \Staest = \frac{1}{n} \sum_{i=1}^n \prod_{t=1}^T [1-\hesttaxtmle] 
    \end{equation}
    \item Estimate the variance by the following empirical plug-in estimates, for efficient influence function $\IF$ and $D_t(a)$ defined in \cref{eq:IFsurv} and \cref{eq:correction} respectively.
    $$\hat V_{t,a}=\E_n[\hat \IF(t,a)^2] - \frac{1-\paest}{\paest} \mathbb{V}_n[\mathbb{E}_n[\hat{D}_t(a) \mid W]] $$
    \item Construct the $1-\alpha$ level CI by
    $$ \Staest \pm z_{1-\alpha} \frac{\sqrt{\hat V_{t,a}}}{\sqrt{n}} $$
\end{enumerate}
Generalizations to other estimands are detailed in \cref{sec:functionals}.

\end{tcolorbox}

Now we justify our procedure. First, we begin with assumptions:

\begin{enumerate}[label={\bfseries (A\arabic*)}]
\tightlist
\item  \textbf{Convergence of outcome regression:}  \label{ass:misspecifiedH}
The estimator $\hat{h}$ converges to some (possibly misspecified) limit $\tilde h$, among all strata $w \in \Sg$
\[
\|\hat{h} - \tilde h \|_{L^2(P^w)} = o_P(1),
\]
for all treatment arms and time points, where we assume $d < \tilde h < 1 - d$ for $d \in (0,1) $ almost surely.

\item \textbf{Nuisance Estimation of Propensity and Censoring Models:}  \label{ass:nuisance_est}
\begin{itemize}
    \item The censoring distribution is estimated via the Nonparametric Maximum Likelihood Estimation (NPMLE) within each treatment arm.
    \item The propensity score is estimated using a parametric model in covariates, parameterized by $\beta$
\end{itemize}
\item \textbf{Model Class Complexity: } Assume $h$ lies in a function class $\Hs$. We make the following assumptions on the model complexity, depending on if we use cross-fitting:
\begin{itemize}
    \item Single-Sample. $\Hs$ is Donsker 
    \item Cross-fitting. We assume that $||p_A-\hat p_A||_{\Lpw} \cdot ||\hat h -  \tilde h||_{\Lpw} = o(n^{-1/2}) $
\end{itemize}
\label{ass:modelComplexity}

\item \textbf{Regularity of TMLE:}  \label{ass:regularity}
Standard regularity conditions for maximum likelihood estimation hold for the logistic regression models used in the TMLE updating steps, as in Theorem 5.7 of \citet{VDV98} s.t.
$$ |\hat \varepsilon - \tilde \varepsilon|_{\Lpw} \rightarrow 0 $$

\end{enumerate}

With these assumptions, we are fit to state our main result which provides inferential guarantees for the survival function under stratified randomization.

\begin{theorem}[Statistical Inference for the Survival Function under Stratified Randomization]\label{thm:surv}
    Assume \labelcref{ass:consistency}-\labelcref{ass:censoring}, and \labelcref{ass:misspecifiedH}-\labelcref{ass:regularity}.
    For all $t \in [T]$, $a \in \{ 0, 1 \}$, we have
    \begin{align*}
        \sqrt{n}(\Staest - \Sta) \rightsquigarrow \N(0, V_{t,a})
    \end{align*}
    for $V_{t,a}=\E[\IF(t,a)^2] - \frac{1-\pa}{\pa} \V[\E[D_t(a) \mid W]] $
\end{theorem}

The proof is found in \labelcref{pf:surv}. Some comments are in order. First, we note that stratification provides a variance refinement, since the variance is the difference between the simple randomization variance and a positive term. This is analogous to Theorem 4.1 of \citet{Wang19}.
\labelcref{ass:sampling} is standard in stratified randomization results and found in \citet{Bugni18, Wang19} among others. \labelcref{ass:censoring} may seem strong, but this assumption is not uncommon in literature (e.g. see Assumption 1' of Theorem 2 from \citep{Wang19} and A1 of Theorem 1 from \citep{Williams22}) and study design choices can be made to mitigate violations to this. \labelcref{ass:misspecifiedH} formalizes that we can attain asymptotic normality even if we have a misspecified outcome regression, although we will not be efficient if misspecified. The proof of asymptotic normality reveals a misspecification term, which is 0 if the outcome model is correct, or treatment and censoring models are correct. This is guaranteed using NPMLE without covariates, as in \labelcref{ass:nuisance_est}. \labelcref{ass:nuisance_est} will permit us to adjust for covariates as we wish in the propensity score model. In contrast to \citet{Williams22}, we do not require strict independence from covariates in the model. \labelcref{ass:modelComplexity} says that we may use sufficiently regular models or data-adaptive, machine-learning based estimators such as lasso or random forest with cross-fitting.

\subsection{Functionals of the Survival Curve}\label{sec:functionals}
We now extend estimation and inference to time-to-event estimands which are smooth functionals of the survival curve. We first begin with examples of interest:
\begin{enumerate}
    \item Risk Difference (RD)
    \begin{equation}\label{eq:RD}
        \RD(t) = \St{1} - \St{0}
    \end{equation}
    \item Restricted Mean Survival Time (RMST)
    \begin{equation}\label{eq:RMST}
        \RMST(t) = \int_0^t \Surv{t'}{1} - \Surv{t'}{0} dF(t')
    \end{equation}
    \item Risk Ratio (RR)
    \begin{equation}\label{eq:RR}
        \RR(t) = \frac{\St{1}}{\St{0}}
    \end{equation}    
    \item Odds Ratio (\OR)
    \begin{equation}\label{eq:OR}
        \OR(t) = \frac{\dfrac{S_1(t)}{1 - S_1(t)}}
     {\dfrac{S_0(t)}{1 - S_0(t)}}
    \end{equation}
    \item Win Ratio (WR)
    \begin{equation}\label{eq:WR}
        \WR(t) = \frac{\pr(T_1 > T_0)}{\pr(T_0 > T_1)}
    \end{equation}
\end{enumerate}

\paragraph{Estimation} To carry out estimation, we plug in the estimate of $\Sta$, $\Staest$ from \cref{eq:estSurv}, and utilize the plug-in estimate for the functional of interest. For example, the RMST will be approximated as $$ \sum_{t=1}^T \hat{S}(t,a) dt$$ and the Win Probability as $$ \sum_{t=1}^T \hat{S}(t, 1) \hat{S}(t, 0) \hat{h}(U_i \geq t \mid \delta_t=1, A=0) $$ (as $dF(t,0) = P(T_0 = t) = P(T_0 = t \mid T_0 \geq t) P(T_0 \geq t)$).

\paragraph{Variance Estimation}
To get variance estimates, we can plug-in $\Staest$ into the EIF, accounting for the reduced variance by $\frac{1}{n} g'(\hat{S}(t,a))^2 \hat{V}_{t,a}$. The definitions and respective influence functions are provided in \cref{tab:functional_estimands}; these follow from the product rule for influence functions.

We justify our proposal with the following corollary: 
\begin{corollary}\label{thm:survEstimand}
    Let $g(\Sta)$ be a functional of $\Sta$ the survival function that is Hadamard differentiable (at its parameters). Assume the conditions of \cref{thm:surv}. Then, we can apply the functional delta method, Theorem 20.8 of \cite{VDV98} to conclude
    \begin{align*}
        \sqrt{n}(g(\hat{S}(t,a))-g(\Sta)) \rightsquigarrow \N(0, V^g_{t,a})
    \end{align*}
    where $V^g_{t,a}= g'(\Sta)$.
\end{corollary}

\begin{table}[ht]
\centering
\renewcommand{\arraystretch}{2}
\begin{tabular}{l l l l}
\hline
Estimand & Definition & Influence Function &  \\
\hline

Survival ($\Sta$)
& $\Sta$ 
& $\phi(t, a)$ 
&  \\

Risk Difference ($\RD$)
& $\St{1} - \St{0}$ 
& $\phi(t,1) - \phi(t,0)$ 
& \\

RMST $(\RMST)$
& $\int_0^t \Sta\,dt$ 
& $\int_0^t (\phi(t',1) - \phi(t',0))dt'$ 
& \\

Risk Ratio (\RR) 
& $\dfrac{\St{1}}{\St{0}}$ 
& $\dfrac{1}{\St{0}}\phi(t,1) - \dfrac{\St{1}}{\St{0}^2}\phi(t,0)$ 
& \\

Odds Ratio (\OR) 
& $\dfrac{S_1(t)}{1 - S_1(t)} / \dfrac{S_0(t)}{1 - S_0(t)}$ 
& $\frac{1-\St{0}}{\St{0}(1-\St{1})^2} \phi(t,1) - \frac{\St{1}}{\St{0}^2(1-\St{1})} \phi(t,0) $ 
& \\

Win Ratio (\WR) 
& $\dfrac{W}{L}$ 
& $\dfrac{1}{L}\phi_{win} - \dfrac{W}{L^2}\phi_{loss}$ 
& \\

Win Probability ($W$)
& $\pr[T_1 > T_0] = \int S(t,1) dF(t, 0)$
& $\int \phi(t,1) dF(t,0) + \int S(t,1) d\phi(t,0)$
& \\

Loss Probability ($L$)
& $\pr[T_0 > T_1] = \int S(t,0) dF(t,1)$
& $\int \phi(t,0) dF(t,1) + \int S(t,0) d\phi(t,1)$
& \\

\hline
\end{tabular}
\caption{Common survival-based estimands, definitions, and influence functions under simple randomization.\label{tab:functional_estimands}}
\end{table}

\clearpage

\section{Simulations} \label{sec:simulations}
In this section, we carry out a simulation study to verify our theoretical properties. Our aim is to demonstrate variance gains under (i) data-adaptive estimation and covariate selection, and (ii) stratified randomization.

\subsection*{Data Generating Mechanism}

For each simulated dataset of size $n$, we generated observed data 
\[
O = (X, A, U, \delta),
\]
where $X$ denotes baseline covariates, $W \subseteq X$ is a stratification variable, $A$ is a binary treatment indicator, $U$ is the observed time, and $\delta$ is the event indicator.  We simulate strata $W$ with varying treatment probabilities and 30 baseline covariates from the uniform distribution (with some variations). Treatment is randomized among strata in a 1:1 ratio. Censoring and event times occur with increasing probability in time. In the event time model, a subset of the odd-indexed covariates yield signal for our event of interest. The estimand of interest is the risk difference at time $3$, or $\theta = S(3,1)-S(3,0)$. 
We detail our simulation study details closer in \cref{app:simulations}.

\subsection*{Results}

The results over 1000 Monte Carlo simulations are depicted in \cref{fig:sim_lasso}, and \cref{fig:sim_rf}. These results depict findings using TMLE with lasso as the nuisance estimator and random forests respectively. Our findings support our underlying theory by empirically demonstrating variance reduction or efficiency gains from our two sources of interest: (i) correct specification of the model with the ability to adapt reasonably well under a data-adaptive approach and (ii) stratification reductions.

\subsection*{Lasso Results}
We begin by considering the bias of our estimators in the top, left panel. Note that the point estimates for simple and stratified cases are the same.  
We see that the incorrectly specified model yields the greatest bias (light blue), followed by our unadjusted estimator the Kaplan-Meier (KM) (black). The model with all covariates and correctly specified covariates perform similarly at larger samples, although the correctly specified model is more performant at smaller sample size ($n=100$). 

In the top, right panel we see the coverage of our estimators. We see that the coverage level is attained for all of our estimators, except for the incorrectly specified model at small sample size ($n=100$). The stratified versions (dashed lines) have marginally tighter widths.

We now consider the relative efficiency with respect to the unadjusted, KM estimator that assumes simple randomization. The plots are shown in the bottom panel. The KM has the lowest efficiency, and the incorrectly specified model is less efficient than its counterparts. Under misspecification, stratification adjustment helps the precision, reducing the variance by approximately 5.4\%. We see that the model with all covariates and correctly specified covariates yield similar variances under simple and stratified randomization assumptions, with marginal gains under stratification. In theory, with correctly specified models, the adjustment term would yield the same variance under simple or stratified randomization supporting this theoretical result.

\subsection*{Random Forest Results}
\begin{figure}[htbp]
\centering
\begin{subfigure}{0.49\textwidth}
  \centering
  \includegraphics[width=\linewidth]{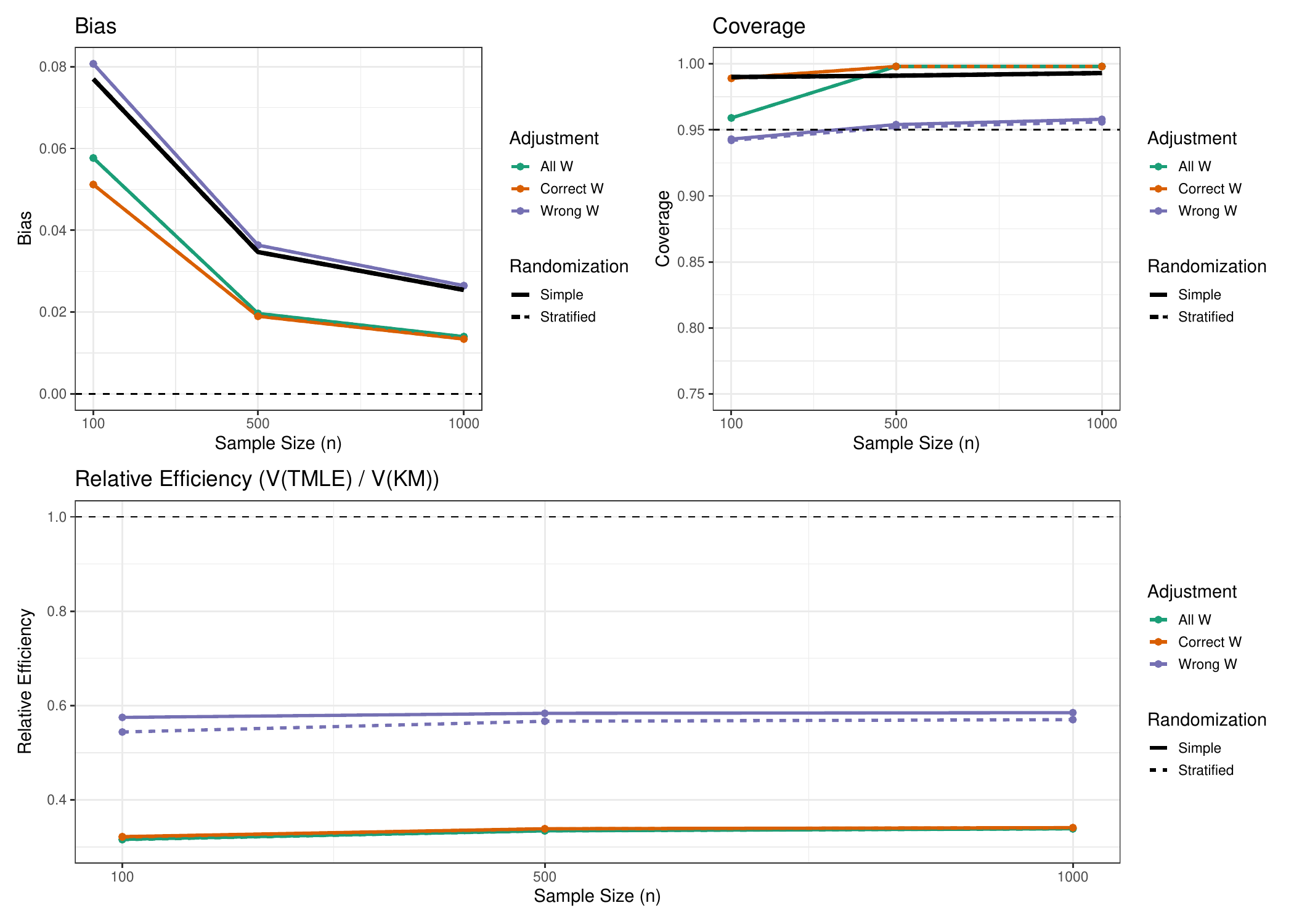}
  \caption{Lasso nuisance estimator}
  \label{fig:sim_lasso}
\end{subfigure}
\hfill
\begin{subfigure}{0.49\textwidth}
  \centering
  \includegraphics[width=\linewidth]{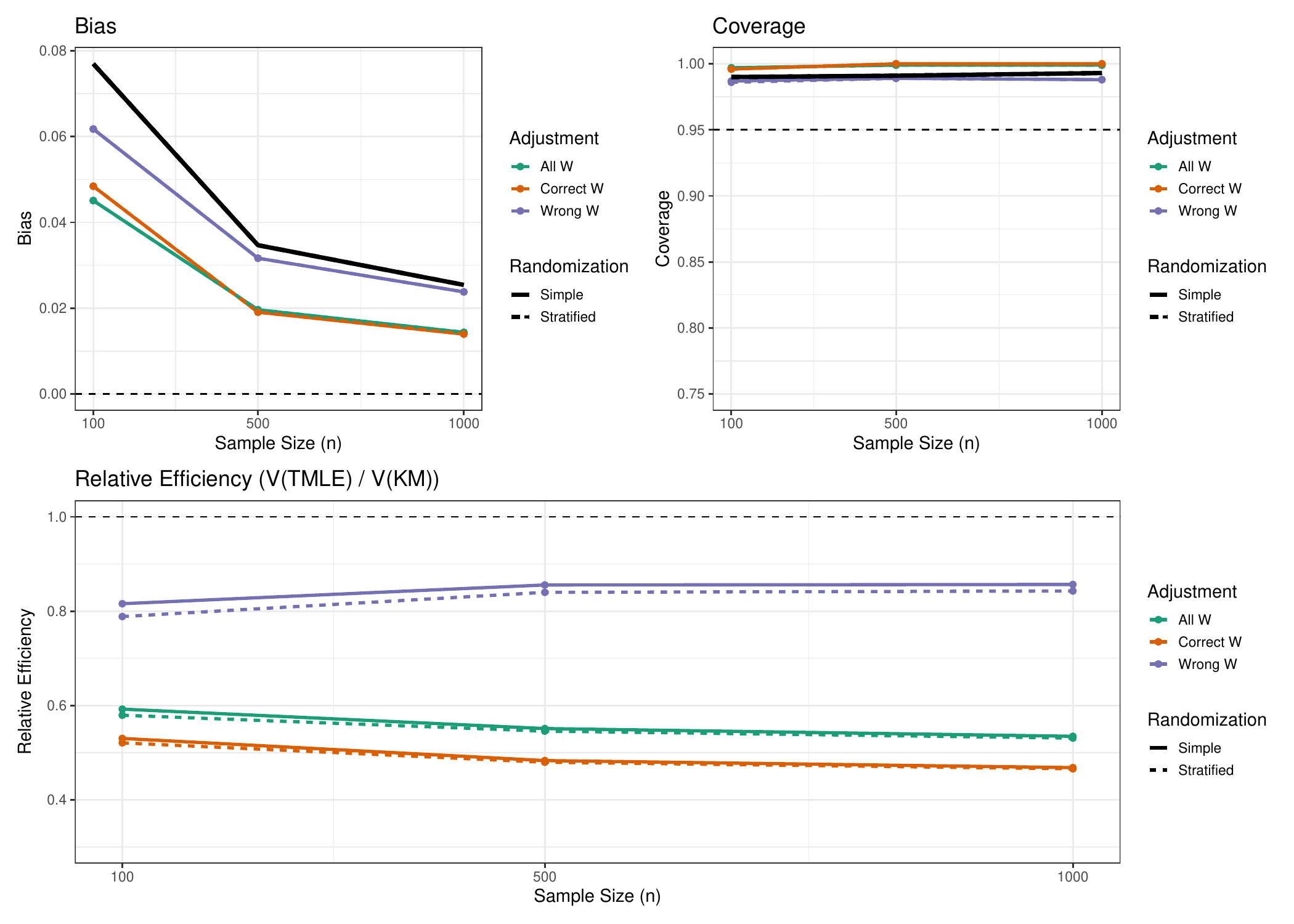}
  \caption{Random Forest nuisance estimator}
  \label{fig:sim_rf}
\end{subfigure}

\caption{Simulation results comparing our proposed estimators to unadjusted estimators over 1000 repetitions.}
\label{fig:sim_combined}
\end{figure}

Similar findings result using random forest as the nuisance estimator, though using this more expressive model demonstrates a bias/variance tradeoff \cref{fig:sim_rf}. Utilizing all covariates leads to lower bias in comparison to the correctly specified model, but higher variance (except for small sample sizes, or $n=100$). The greatest precision gains are seen with the stratified, correctly specified random forest model. This model finds greater precision gains from stratified adjustments compared to the more correct linear specification from lasso, aligning with our theory.  It is also interesting to note that the incorrectly specified model has increasing variance as sample sizes increase.
We finally remark that the less expressive lasso model, compared to the random forest, yields higher bias but lower variance.

\section{Data Application}\label{sec:rwd}

Randomized clinical trials with small samples sizes are not atypical depending on the therapeutic area or logistical constraints. An unfortunate consequence of this is the decreased precision and power. In this section, we illustrate the usability of our methods by reanalyzing the DECIFER trial \citep{Goff19}, an RCT evaluating the effect of citalopram on patients with first-episode schizophrenia (FES). The treatment of interest was citalopram vs. control. The outcome of interest was time to first improvement of $M$ in Calgary Depression Scale for Schizophrenia (CDSS). The estimand of interest was taken to be the risk difference at 52 weeks for time to first improvement of $M$ in CDSS score, between those on citalopram and control. We carry out our primary analysis on the median ($M=-1$), and explore sensitivity of our findings to choosing $M$ to be the mean ($M=-1.65$) or 25th percentile ($M=-2$).

\subsection*{Preprocessing }
The DECIFER trial enrolled 95 participants, with 52 completing the 12-month assessment. We constructed the analysis dataset from the DECIFER trial by excluding individuals with missing treatment or outcome data and selecting baseline covariates from demographics, psychiatric history, clinical surveys, and imaging. Two samples are removed since they have the covariate duration of untreated psychosis missing, a critical covariate for this trial \citep{Goff19}. After preprocessing, described in greater details in \cref{sec:preprocessing}, we find a final analytical sample of $n=50, p = 34$. The covariates, treatment, and outcomes used are found in \cref{tab:covariates}.

\subsection*{Results}
\begin{figure}[ht!]
\includegraphics[width=1\textwidth]{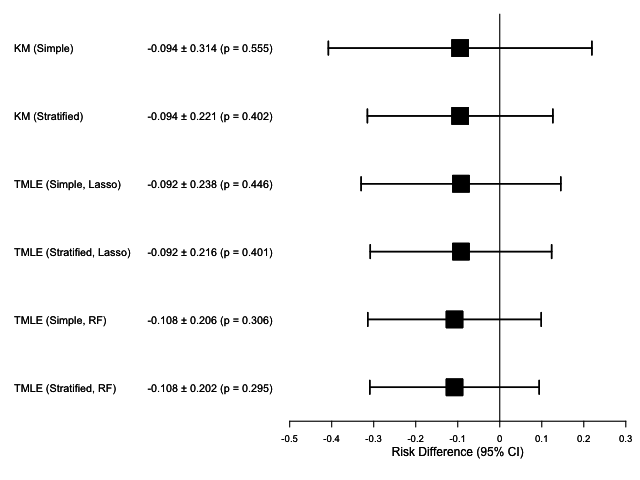}
\caption{Forest Plot of our analyses, with a median threshold change. \label{fig:median}}
\end{figure}

Our findings are displayed in the forest plot on the median change \cref{fig:median}. We run the unadjusted KM estimator, covariate adjustment method TMLE with Lasso, and covariate adjustment method TMLE with Random Forests (RF). We examine the variances under simple and stratified randomization assumptions.

Overall, the point estimates across the methods are quite similar, suggesting that the treatment decreases the time to first improvement of $M$ in CDSS. However, the variances differ across the methods, demonstrating gains from covariate adjustment and stratification.

Covariate adjustment substantially improved efficiency relative to the unadjusted KM estimator. Under simple randomization, covariate adjustment (TMLE) reduces variance by approximately 42-56\%. Under stratification, the additional benefit of covariate adjustment is present, but attenuated, providing a more modest 4.4-16.6\% variance reduction.

Comparing stratified to simple randomization, the lasso-based estimator achieves a notable variance reduction of approximately 17.6\%, while the random forest estimator exhibits 4.7\%. The gains for the unadjusted estimator are notable.

\begin{table}
\begin{tabular}[t]{llllrr}
\toprule
Method & Randomization & Model & Estimate (95\% CI) & Variance & p-value\\
\midrule
KM & Simple & -- & -0.094 (-0.408, 0.219) & 1.280 & 0.555\\
KM & Stratified & -- & -0.094 (-0.315, 0.126) & 0.634 & 0.402\\
\addlinespace
TMLE & Simple & Lasso & -0.092 (-0.330, 0.145) & 0.735 & 0.446\\
TMLE & Stratified & Lasso & -0.092 (-0.308, 0.123) & 0.606 & 0.401\\
\addlinespace
TMLE & Simple & RF & -0.108 (-0.314, 0.099) & 0.555 & 0.306\\
TMLE & Stratified & RF & -0.108 (-0.309, 0.094) & 0.529 & 0.295\\
\bottomrule
\end{tabular} 
\caption{Table with DECIFER application results \label{tab:rwd}}
\end{table}

\section{Discussion}\label{sec:conclusion}
Time-to-event outcomes are central to many randomized clinical trials, yet methods that fully integrate stratified randomization with flexible, data-adaptive covariate adjustment remain underdeveloped. In this work, we proposed targeted minimum loss-based estimators (TMLE) that accommodate general time-to-event estimands (any smooth functional of the survival curve) while leveraging both stratification and modern machine learning–based adjustment. 
This flexible yet general framework is particularly valuable in modern clinical trials, where high-dimensional baseline data are routinely collected. 
Our theoretical results and analyses demonstrate that precision gains can be achieved from two complementary sources: covariate adjustment with data-adaptive estimators and stratified randomization.

From a practical standpoint, our methods are straightforward to implement. Existing TMLE software can be used with minimal modification, requiring only a correction to the variance estimator to account for stratified designs. This lowers the barrier to adoption and facilitates application in real-world trial settings. Moreover, our framework extends naturally to other outcome types, allowing practitioners to apply a unified approach across a variety of endpoints. Our additional simulations and empirical analyses, found in \cref{sec:tmle_generalizations}, with non-time-to-event outcomes support the efficiency gains possible.

We note a couple of limitations. First, while our analysis relies on standard assumptions, certain assumptions may be violated in practice such as independent censoring, though steps can be taken to mitigate this possibility. Second, we develop our methods based on discrete time settings which is common in many RCTs. When working with continuous time, we recommend discretizing the time, which could introduce bias. 
Future work could extend our methods to settings with informative censoring, or more complex randomization schemes.

In summary, this work contributes to the growing literature on covariate adjustment in randomized trials by providing a flexible and principled approach for analyzing time-to-event outcomes under stratified randomization. By combining modern machine learning with rigorous causal inference methodology, our approach offers a practical pathway toward efficient and interpretable analyses in clinical trials.

\bibliographystyle{plainnat}
\bibliography{paper-ref}

@article{Bannick25,
  author = {Bannick, M. S. and Shao, J. and Liu, J. and Du, Y. and Yi, Y. and Ye, T.},
  title = {A general form of covariate adjustment in clinical trials under covariate-adaptive randomization},
  journal = {Biometrika},
  volume = {112},
  number = {3},
  pages = {asaf029},
  year = {2025}
}

@book{Billingsley,
  author = {Billingsley, P.},
  title = {Convergence of Probability Measures},
  publisher = {John Wiley \& Sons},
  address = {New York},
  series = {Wiley Series in Probability and Statistics},
  edition = {2},
  year = {1999}
}

@article{Bugni18,
  author = {Bugni, F. A. and Canay, I. A. and Shaikh, A. M.},
  title = {Inference under covariate-adaptive randomization},
  journal = {Journal of the American Statistical Association},
  volume = {113},
  number = {524},
  pages = {1784--1796},
  year = {2018}
}

@book{Durrett,
  author = {Durrett, R.},
  title = {Probability: Theory and Examples},
  publisher = {Cambridge University Press},
  address = {Cambridge},
  series = {Cambridge Series in Statistical and Probabilistic Mathematics},
  edition = {4},
  year = {2019}
}

@article{FayLi24,
  author = {Fay, M. P. and Li, F.},
  title = {Causal interpretation of the hazard ratio in randomized clinical trials},
  journal = {Clinical Trials},
  volume = {21},
  number = {5},
  pages = {623--635},
  year = {2024}
}

@article{Goff19,
  author = {Goff, D. C. and Freudenreich, O. and Cather, C. and Holt, D. and Bello, I. and Diminich, E. and Tang, Y. and Ardekani, B. A. and Worthington, M. and Zeng, B. and Wu, R. and Fan, X. and Li, C. and Troxel, A. and Wang, J. and Zhao, J.},
  title = {Citalopram in first episode schizophrenia: The DECIFER trial},
  journal = {Schizophrenia Research},
  volume = {208},
  pages = {331--337},
  year = {2019}
}

@book{KleinbaumKlein12,
  author = {Kleinbaum, D. G. and Klein, M.},
  title = {Survival Analysis: A Self-Learning Text},
  publisher = {Springer},
  address = {New York},
  edition = {3},
  year = {2012}
}

@article{Rafi23,
  author = {Rafi, A.},
  title = {Efficient semiparametric estimation of average treatment effects under covariate adaptive randomization},
  journal = {arXiv preprint arXiv:2305.08340},
  year = {2023}
}

@article{Stensrud19,
  author = {Stensrud, M. J. and Aalen, J. M. and Aalen, O. O. and Valberg, M.},
  title = {Limitations of hazard ratios in clinical trials},
  journal = {European Heart Journal},
  volume = {40},
  number = {17},
  pages = {1378--1383},
  year = {2019}
}

@book{Tsiatis,
  author = {Tsiatis, A. A.},
  title = {Semiparametric Theory and Missing Data},
  publisher = {Springer},
  address = {New York},
  series = {Springer Series in Statistics},
  year = {2006}
}

@article{Tsiatis08,
  author = {Tsiatis, A. A. and Davidian, M. and Zhang, M. and Lu, X.},
  title = {Covariate adjustment for two-sample treatment comparisons in randomized clinical trials: A principled yet flexible approach},
  journal = {Biometrics},
  volume = {64},
  number = {3},
  pages = {650--659},
  year = {2008}
}

@article{Tu24,
  author = {Tu, F. and Ma, W. and Liu, H.},
  title = {A unified framework for covariate adjustment under stratified randomisation},
  journal = {Stat},
  volume = {13},
  number = {4},
  pages = {e70016},
  year = {2024}
}

@book{VDVW96,
  author = {van der Vaart, A. and Wellner, J. A.},
  title = {Weak Convergence and Empirical Processes: With Applications to Statistics},
  publisher = {Springer},
  address = {New York},
  year = {1996}
}

@book{VDV98,
  author = {van der Vaart, A. W.},
  title = {Asymptotic Statistics},
  publisher = {Cambridge University Press},
  address = {Cambridge},
  year = {1998}
}

@article{VLDV24,
  author = {Van Lancker, K. and D{\'i}az, I. and Vansteelandt, S.},
  title = {Automated, efficient and model-free inference for randomized clinical trials via data-driven covariate adjustment},
  journal = {arXiv preprint arXiv:2404.11150},
  year = {2024}
}

@article{Wang19,
  author = {Wang, B. and Susukida, R. and Mojtabai, R. and Amin-Esmaeili, M. and Rosenblum, M.},
  title = {Model-robust inference for clinical trials that improve precision by stratified randomization and covariate adjustment},
  journal = {Journal of the American Statistical Association},
  volume = {118},
  number = {542},
  pages = {1152--1163},
  year = {2021}
}

@article{Williams22,
  author = {Williams, N. and Rosenblum, M. and D{\'i}az, I.},
  title = {Optimising precision and power by machine learning in randomised trials with ordinal and time-to-event outcomes with an application to COVID-19},
  journal = {Journal of the Royal Statistical Society Series A: Statistics in Society},
  volume = {185},
  number = {4},
  pages = {2156--2178},
  year = {2022}
}

\clearpage 

\appendix

\renewcommand{\thesection}{\Alph{section}}

\begin{center}
    \textbf{\large Supplementary Materials to\\ ``\titleString''}\\ \vspace{0.25cm}
\end{center}

\section{Proofs}
We first establish helper results.
\subsection{Helper Quantities and Lemmas}

\textbf{Influence Function}
The efficient influence function for $\Sta$ is given by
\begin{equation}\label{eq:IFsurv}
\phi_{t,a} = \sum_{t=1}^T I_t \left( - \frac{\mathbf{1}(A=a)}{\pa(X)} \frac{S(T,a,X)}{S(t,a,X)}\frac{1}{\Pi_C(t,a,X)} \right) \{L_t - h(t,a,X)\}  + S(t,a,X) - S(t,a).
\end{equation}
for
\[
L_t = \mathbf{1}\{U = t,\, \delta = 1\},  \qquad I_t = \mathbf{1}\{U \ge t\}
\]

\textbf{Correction Term}
The correction term $D$ is given by
\begin{equation}\label{eq:correction}
    D_t(a) = \sum_{t=1}^T I_t \left( -\frac{S(T,a,X)}{S(t,a,X)}\frac{1}{\Pi_C(t,a,X)} \right) \{L_t - h(t,a,X)\}
\end{equation}

\begin{lemma}[Restatement of Lemma 2 of \citet{Wang19}]\label{lemma:lemma2Wang}
With overloading on $Z$, let
\[
Z_i(1) = g_1\big(Y_i(1), C_i(1), X_i\big),
\qquad
Z_i(0) = g_0\big(Y_i(0), C_i(0), X_i\big),
\]
for some measurable functions $h_1$ and $h_2$ such that
\[
\E[Z_i(a)^2] < \infty, \qquad a=0,1.
\]
Then under stratified randomization or the biased-coin design,
\[
\frac{1}{\sqrt{n}} \sum_{i=1}^n \Big\{ A_i\big(Z_i(1)-\E[Z_i(1)]\big) - (1-A_i)\big(Z_i(0)-\E[Z_i(0)]\big)
\Big\}. \xrightarrow{d} \mathcal{N}(0,V_z),
\]
where
\[
V_z = \V\!\big( p_a\{Z(1)-\E[Z(1)]\} - (1-p_a)\{Z(0)-\E[Z(0)]\} \big) + p_a(1-p_a) \V\!\big( \E[Z(1)-Z(0)\mid W] \big).
\]
\end{lemma}

\subsection{Proof of Theorem \texorpdfstring{\labelcref{thm:surv}}{Theorem}}
\label{pf:surv}

Decompose the estimator for target parameter as follows:
\begin{align*}
    \sqrt{n}(\hat{S}(t,a) - \Sta) &= \sqrt{n} \Ptrue \IFta(\cdot; \hat{\eta}) + \sqrt{n} R_{n}(\hat{\eta}) \\
    &= \Gn\IFta(\cdot; \hat{\eta}) + \sqrt{n} R_n(\hat{\eta}) \\
    &= \underbrace{\Gn \IFta(\cdot; \tilde \eta)}_{\mbox{Effective Term}} + \underbrace{\Gn \big[\IFta(\cdot; \hat{\eta})-\IFta(\cdot; \tilde \eta) \big]}_{B_{n,2}} + \underbrace{\sqrt{n} R_n(\hat{\eta})}_{B_{n,1}} 
\end{align*}
for nuisance parameters $\eta=(\Pi_C, p_A, h)$, consisting of the the censoring, treatment, and conditional hazard model. We address these terms one-by-one following \citet{Williams22}, showing asymptotic normality of the first term, and that the bias terms are negligible, but modify arguments to account for stratification and treatment dependence on the covariates.
Recall that $\hat \eta$ is the TMLE solution and $\tilde \eta$ is the misspecified limit.

\subsubsection*{Empirical Process Term: $B_{n,2}$}

\textbf{Donsker case}

We decompose this term as follows:
\begin{align*}
\IF_{t,a}(\cdot; \hat{\eta})-\IF_{t,a}(\cdot; \tilde \eta) &= \big[\IF_{t,a}(\cdot; \hat{h}, \hat{p}_A, \hat{\Pi}_C) - \IF_{t,a}(\cdot; \tilde{h}, \hat{p}_A, \hat{\Pi}_C)\big] \\
&\quad + \big[\IF_{t,a}(\cdot; \tilde{h}, \hat{p}_A, \hat{\Pi}_C) - \IF_{t,a}(\cdot; \tilde{h}, p_A, \hat{\Pi}_C)\big] \\
&\quad + \big[\IF_{t,a}(\cdot; \tilde{h}, p_A, \hat{\Pi}_C) - \IF_{t,a}(\cdot; \tilde{h}, p_A, \Pi_C)\big] \\
&=: \Delta_h + \Delta_{p_A} + \Delta_{\Pi_C}.
\end{align*}

Thus,
\begin{align*}
B_{n,2}
= \mathbb{G}_n(\Delta_h) + \mathbb{G}_n(\Delta_{p_A}) + \mathbb{G}_n(\Delta_{\Pi_C}).
\end{align*}

We know that $\E[\IFta(\cdot ; \hat \eta) - \IFta(\cdot ; \tilde \eta)] = \E[\E[\IFta(\cdot ; \hat \eta) - \IFta(\cdot ; \tilde \eta) \mid W]]$.
By \labelcref{ass:misspecifiedH} and \labelcref{ass:nuisance_est}, conditioning on strata, and invoking the convergence results, each term conditional on $W=w$ is then $\opo$ and independent among $W=w$. Further assuming that $\Hs$ is Donsker, we can apply Theorem 19.24 of \citet{VDV98} to each stratum-specific term yielding $|B_{n,2}| = o_p(1)$. 

\textbf{Cross-fitted case}

With cross-fitting, our empirical process term is given by
\begin{align*}
B_{n,2}  = \frac{1}{\sqrt{J}} \sum_{j=1}^J \mathsf{G}_{n,j} \big( \IF_{t,a}(\cdot ; \hat{\eta}_j) - \IF_{t,a}(\cdot; \tilde \eta) \big),
\end{align*}
where $\hat{\eta}_j = (\hat{h}_j, \hat{p}_{A,j}, \hat{\pi}_{C,j})$ is trained on $\mathcal{T}_j$.

Define
\[
\mathcal{F}_n^j  = \left\{ \IFta(\cdot; \hat{\eta}_j)-\IFta (\cdot; \tilde \eta) :\, |\hat \varepsilon - \tilde \varepsilon| < d_n \right\},
\]
with envelope
\[
F_n^j  = \sup_{|\varepsilon - \tilde \varepsilon| < d_n} \left| \IF_{t,a}(\cdot; \hat{\eta}_j)-\IF_{t,a}(\cdot; \tilde \eta)  \right|.
\]

Because $\hat{\eta}_j$ is fixed given the training data, Theorem 2.14.2 of \cite{VDVW96} yields the following:
\begin{align*}
\E\!\left\{ \sup_{f \in \mathcal{F}_n^j} \big| \mathsf{G}_{n,j} f \big| \,\Big|\, \mathcal{T}_j, \Sn \right\}  \lesssim\; \|F_n^j\| \int_0^1 \sqrt{ 1 + N_{[]}\!\big( \alpha \|F_n^j\|, \mathcal{F}_n^j, L_2(P) \big) } \, d\alpha,
\tag{8}
\end{align*}
Note that we have conditioned on the training fold $\mathcal{T}_j$ \textit{and} the strata $W^{(n)}$ to induce independence. 

Employing Theorem 2.7.2 of \cite{VDVW96}, we find
\[
\log N_{[]}\!\big( \alpha \|F_n^j\|, \mathcal{F}_n^j, L_2(P) \big) \;\lesssim\; \frac{1}{\alpha \|F_n^j\|}.
\]

Hence, 
\begin{align*}
\|F_n^j\| \int_0^1 \sqrt{ 1 + N_{[]}\!\big( \alpha \|F_n^j\|, \mathcal{F}_n^j, L_2(P) \big) } \, d\alpha &\;\lesssim\; \int_0^1 \sqrt{ \|F_n^j\|^2 + \frac{\|F_n^j\|}{\alpha} } \, d\alpha \\ &\;\le\; \|F_n^j\| + \|F_n^j\|^{1/2} \int_0^1 \alpha^{-1/2} \, d\alpha \\ &\;\le\; \|F_n^j\| + 2 \|F_n^j\|^{1/2} \\
&= \opo
\end{align*}
where our last step follows by consistency of $\hat{\IF}_{t,a}$ to $\IFta$ and \labelcref{ass:regularity}, which guarantees we can find a deterministic sequence $d_n \to 0$ such that our fluctuation parameter is within $d_n$ of the possibly misspecified limit.

\subsubsection*{Remainder Term: $B_{n,1}$}

\begin{align*}
B_{n,1} 
&= -  \sqrt{n} \int \sum_{t=1}^T 
    \frac{\hat S(T,a,x)}{\hat S(t,a,x)} S(t-1,a,x) 
    \{\hat h(t,a,x) - h(t,a,x)\} 
    [ \frac{p_A(a,x) \Pi_C(t,a)}{\hat p_A(a,x) \hat \Pi_C(t,a)} -1 ] \, dP(X) \\
&= -  \sqrt{n} \sum_w p_w \int \sum_{t=1}^T 
    \frac{\hat S(T,a,x)}{\hat S(t,a,x)} S(t-1,a,x) 
    \{\hat h(t,a,x) - h(t,a,x)\} 
    [ \frac{p_A(a,x) \Pi_C(t,a)}{\hat p_A(a,x) \hat \Pi_C(t,a)} -1 ] \, dP^w(X) \\
&=:  \sqrt{n} \sum_w p_w B_{n,1}^{(w)}.
\end{align*}

\noindent
We now analyze $B_{n,1}^{(w)}$ utilizing a Taylor expansions to handle the dependence of the treatment propensity on covariates.

\begin{align*}
    \frac{p_A(a,x) \Pi_C(t,a)}{\hat p_A(a,x) \hat \Pi_C(t,a)} - 1 &= \frac{p_A(a,x)}{\hat p_A(a,x)} - 1 +  \frac{\Pi_C(t,a)}{\hat \Pi_C(t,a)}  - 1 + (\frac{p_A(a,x)}{\hat p_A(a,x)} - 1 )(\frac{\Pi_C(t,a)}{\hat \Pi_C(t,a)}  - 1)  \\
    &= \frac{(p_A(a,x)-\hat p_A(a,x))}{p_A(a,x)} + \frac{(\hat \Pi_C(t,a) - \Pi_C(t,a))}{\Pi_C(t,a)} + O(||\hat p_A - p_A|| \cdot || \hat \Pi_C - \Pi_C || ) \\
    &= - \frac{(\hat \beta - \beta) \partial \beta [p_A(a,x)]}{p_A(a,x)}  + \frac{(\hat \Pi_C(t,a) - \Pi_C(t,a))}{\Pi_C(t,a)} + O(||\hat p_A - p_A|| \cdot || \hat \Pi_C - \Pi_C || )  \\
\end{align*}
\noindent
Plugging this into $B_{n,1}^{(w)}$ gives
\begin{align*}
B_{n,1}^{(w)} 
&= -\sqrt{n}  \int \sum_{t=1}^T 
   \frac{\hat S(T,a,x)}{\hat S(t,a,x)} S(t-1,a,x) 
   \{\hat h(t,a,x) - h(t,a,x)\} \frac{(\hat \beta - \beta) \partial \beta [p_A(a,x)]}{p_A(a,x)} \, dP^w(x) \\
& \quad +  \sqrt{n}  \int \sum_{t=1}^T 
   \frac{\hat S(T,a,x)}{\hat S(t,a,x)} S(t-1,a,x) 
   \{\hat h(t,a,x) - h(t,a,x)\} \frac{(\hat \Pi_C(t,a) - \Pi_C(t,a))}{\Pi_C(t,a)} \, dP^w(x) + o_P(1) \\
&= -\sqrt{n} (\hat \beta - \beta) \int\frac{\partial \beta [p_A(a,x)]}{p_A(a,x)} \sum_{t=1}^T 
   \frac{\hat S(T,a,x)}{\hat S(t,a,x)} S(t-1,a,x) 
   \{\hat h(t,a,x) - h(t,a,x)\}  \, dP^w(x) \\
& \quad +  \sqrt{n} \frac{(\hat \Pi_C(t,a) - \Pi_C(t,a))}{\Pi_C(t,a)}   \int \sum_{t=1}^T 
   \frac{\hat S(T,a,x)}{\hat S(t,a,x)} S(t-1,a,x) 
   \{\hat h(t,a,x) - h(t,a,x)\}  \, dP^w(x) + o_P(1) \\
&= -\underbrace{\sqrt{n} (\hat \beta - \beta) \int\frac{\partial \beta [p_A(a,x)]}{p_A(a,x)} (S(T,a,x) - \hat S(T,a,x)) \, dP^w(x)}_{B_{n,1}^{(w, \beta)}} \\
& \quad +  \underbrace{\sqrt{n} \frac{(\hat \Pi_C(t,a) - \Pi_C(t,a))}{\Pi_C(t,a)}   \int (S(T,a,x) - \hat S(T,a,x))  \, dP^w(x)}_{B_{n,1}^{(w, C)}} + o_P(1) 
\end{align*}
where the last step follows by Lemma 2 of \citet{Williams22}. Now we handle each of these more closely.

\begin{align*}
B_{n,1}^{(w, \beta)} &= \sqrt{n} (\hat \beta - \beta) \int\frac{\partial \beta [p_A(a,x)]}{p_A(a,x)} (S(T,a,x) - \hat S(T,a,x)) \, dP^w(x) \\
 & \leq \sqrt{n} (\hat \beta - \beta) \sqrt{\int\frac{\partial \beta [p_A(a,x)]}{p_A(a,x)} dP^w(x)} \sqrt{\int (S(T,a,x) - \hat S(T,a,x)) \, dP^w(x)} \quad \mbox { by C.S.}\\
 &= \Opo \cdot \opo
\end{align*}
where the last line follows by MLE regularity conditions on the propensity score along with boundedness of the norm score, and consistency of the survival estimate. Consistency of the survival estimate is clear noting the doubly-robust consistency property; $\IFta$ is mean 0 conditional on $\Sn$ (so long as $\hat h$ or $\hat{p}_a$ and $\hat{p}_c$ are correctly specified; in fact, this is guaranteed by the NPMLE \labelcref{ass:nuisance_est}).

Now the second term is handled in a similar manner, without the extra complication of propensity score dependencies. Namely, asymptotic normality of the censoring distribution (an NPMLE estimator with no covariate dependence is trivial). This yields
\begin{align*}
    B_{n,1}^{(w, C)} &= \sqrt{n} \frac{(\hat \Pi_C(t,a) - \Pi_C(t,a))}{\Pi_C(t,a)}   \int (S(T,a,x) - \hat S(T,a,x))  \, dP^w(x) \\
    & = \Opo \opo
\end{align*}

\subsubsection*{Effective Term}

By the influence function, we have:
\begin{align*}
    \frac{1}{\sqrt{n}} \sum_{i=1}^n \sum_{t=1}^T I_t \left( - \frac{\mathbf{1}(A=a)}{\pa(X)} \frac{S(T,a,X)}{S(t,a,X)}\frac{1}{\Pi_C(t,a,X)} \right) \{L_t - \tilde h(t,a,X)\}  + S(t,a,X) - S(t,a) 
\end{align*}

We consider the first term of this expression (the weighted residual), and the second term (difference between the conditional survival and marginal survival on covariates) separately.
\newline\newline
\textbf{First term (weighted residual).}
Define $D_t(a) = \sum_{t=1}^T I_t \left( -\frac{S(T,a,X)}{S(t,a,X)}\frac{1}{\Pi_C(t,a,X)} \right) \{L_t - \tilde h(t,a,X)\}$.

Then, analyze the variance of this expression as
\begin{align*}
    \frac{1}{\sqrt{n}} & \sum_{i=1}^n \sum_{t=1}^T I_t \left( - \frac{\mathbf{1}(A=a)}{\pa(X)} \frac{S(T,a,X)}{S(t,a,X)}\frac{1}{\Pi_C(t,a,X)} \right) \{L_t - \tilde h(t,a,X)\} \\
    & \qt = \frac{1}{\sqrt{n}} \sum_{i=1}^n \frac{\mathbf{1}(A=a)}{\pa(X)}  \sum_{t=1}^T I_t \left( -\frac{S(T,a,X)}{S(t,a,X)}\frac{1}{\Pi_C(t,a,X)} \right) \{L_t - \tilde h(t,a,X)\}  \\
    & \qt = \frac{1}{\sqrt{n}} \sum_{i=1}^n \frac{\mathbf{1}(A=a)}{\pa(X)} D_t(a) \\
    & \qt \rightsquigarrow \N(0, \sigma_{1,a}^2 + \sigma_{2,a}^2)
\end{align*}
where the last step follows by Lemma 2 of \citet{Wang19}  (restated below in \cref{lemma:lemma2Wang}). Above, the variance terms are given by
\[
\sigma_{1,a}^2=\frac{1}{\pa} \V\!\big( D_t(a) - \E[D_t(a) \mid W] \big),
\]

\[
\sigma_{2,a}^2 = \V\!\big( \E[D_{t}(a)\mid W] \big).
\]
for $D_t(a)$ defined in \cref{eq:correction}. 

\textbf{Second term (Difference between conditional and marginal survival).}

The second term satisfies:
\[
\frac{1}{\sqrt{n}} \sum_{i=1}^n \big( S(t,a,X_i) - S(t,a) \big) \xrightarrow{d} \mathcal N\big(0, \V(S(t,a,X))\big),
\]
since $\{X_i\}$ are i.i.d. and independent of the treatment balancing mechanism.
\newline\newline
\subsubsection*{Final Asymptotic Distribution}

Therefore,
\[
\sqrt{n}(\hat S(t,a) - S(t,a)) \rightsquigarrow \mathcal N(0, \Vta),
\]
where
\begin{align*}
    \Vta &= \sigma_{1,a}^2 + \sigma_{2,a}^2 + \V(S(t,a,X)) \\
    &= \frac{1}{\pa} \V\!\big( D_t(a) - \E[D_t(a) \mid W] \big) + \V\!\big( \E[D_{t}(a)\mid W] \big) + \V(S(t,a,X))
\end{align*}

\subsubsection*{Variance Reduction under Stratified Randomization}

We can rewrite this variance expression to demonstrate the variance gain from stratified randomization
\begin{align*}
    V_{t,a} &= \sigma_{1,a}^2 + \sigma_{2,a}^2 + \V(S(t,a,X)) \\
    &= \frac{1}{\pa} \V\!\big( D_t(a) - \E[D_t(a) \mid W] \big) + \V\!\big( \E[D_{t}(a)\mid W] \big) + \V(S(t,a,X)) \\
    \V[\IFta^2] &= \frac{1}{\pa} \V[D_t(a)] + \V[S(t,a,X)] \\
    &= \frac{1}{\pa} [\V[D_t(a) - \E[D_t(a) \mid W]] + \V[\E[D_t(a) \mid W]]] + \V[S(t,a,X)] 
\end{align*}

Hence, the variance reduction is given by
\begin{align*}
    V_{t,a} - \V[\IFta^2] &= (\frac{1}{\pa} - 1) \V[\E[D_t(a) \mid W]] \\
    &= \frac{1-\pa}{\pa} \V[\E[D_t(a) \mid W]] \\
    & > 0
\end{align*}

\subsubsection*{Consistent Variance Estimation}

Consistent variance estimation follows by plugging into the empirical form of our estimator. We demonstrate this on the correction term; the full estimation generalizes in a similar manner.

Note that we have consistency of strata to stratum probabilities $p_s$ by the LLN. Define 
\begin{align*}
    \hat D_{i,t}^a &= \sum_{t=1}^T I_t \left( -\frac{\hat S(T,a,X)}{\hat S(t,a,X)}\frac{1}{\hat{\Pi}_C(t,a,X)} \right) \{L_t - \hat h(t,a,X)\}
\end{align*}
Since we have consistent estimation of our quantities (to some possibly misspecified limit), it suffices to decompose our expression by arms and demonstrate that the true quantities converge to its mean.
\begin{align*}
    \frac{1}{n} \sum_{i=1}^n (1-\I{A_i=a}) D_{i,t}^0 + \I{A_i=a} D_{i,t}^0 \rightarrow \E[D_t(a) \mid W] 
\end{align*}

In fact, if our quantities are cross-fit, this follows by part (1) of Lemma 1 of \citet{Wang19}. If these quantities are Donsker, then we utilize part (2) of the same Lemma.

\section{Additional Simulation Details}\label{app:simulations}
\paragraph{Stratification variable.}
We generated $w \in \{1,2,3,4\}$ from a categorical distribution with probabilities proportional to the stratum index:
\[
P(S = s) = \frac{s}{\sum_{j=1}^4 j}.
\]

\paragraph{Baseline covariates.}
We generated $W = (W_1, \dots, W_{30})$ as follows. 
For odd indices, $W_1, W_3, W_5, W_7, W_9, W_{11}, W_{13}, W_{15}$ are drawn from normal distributions, while the remaining variables are drawn from uniform distributions. Specifically,
\begin{itemize}
\item $W_1 \sim N(0.125,1)$, \quad $W_3 \sim N(- 0.125,1) $, \quad $W_5 \sim N(0.05,1)$,
\item $W_7, W_9, W_{15} \sim N(0,1)$,
\item $W_{11}, W_{13} \sim N(0, 1/16)$,
\item all remaining variables $W_j$ are independently drawn from $\mathrm{Unif}[-1,1]$.
\end{itemize}

\paragraph{Treatment assignment.}
Treatment was assigned via stratified randomization with equal allocation within each stratum:
\[
A \mid W \sim \text{Bernoulli}(0.5)
\]

\paragraph{Event time.}
Event times were generated in discrete time $t = 1, \dots, t_{\max}$ using a pooled logistic model. A subset of the odd indexed covariates were signal variables, while the remaining covariates were noise. This construction yields a mix of prognostic signal and irrelevant features. The hazard at time $t$ is given by
\begin{align*}
\mbox{logit}\{ h_T(t) \}
&= \frac{-3.1 + 0.55\,t}{5} \\
&\quad + 1.5\,\mathbb{I}(S=2)
- 1.5\,\mathbb{I}(S=3)
+ 1.5\,\mathbb{I}(S=4) \\
&\quad - 0.2\,A
- 0.0005\,tA \\
&\quad + 0.03\,t\,\mathbb{I}(S=3)
+ 0.03\,t\,\mathbb{I}(S=4) \\
&\quad - 0.025\,A\,\mathbb{I}(S=3)
- 0.070\,A\,\mathbb{I}(S=4) \\
&\quad - 0.009\,tA\,\mathbb{I}(S=4) \\
&\quad + 3\,W_1 + 3\,W_3 + 3\,W_5
\end{align*}

This yields a realistic increase in events over time, with treated groups having lower event rates.

\begin{table}[ht!]
\centering
\begin{tabular}{c|cc}
\hline
$t$ & $P(\text{event} \mid A=0)$ & $P(\text{event} \mid A=1)$ \\
\hline
1 & 0.09 & 0.08 \\
2 & 0.10 & 0.09 \\
3 & 0.11 & 0.10 \\
4 & 0.13 & 0.11 \\
\hline
\end{tabular}
\caption{Approximate marginal event hazards by treatment over time under the DGM.}
\end{table}


\paragraph{Censoring time.}
Censoring times were generated from a discrete-time hazard:
\[
\text{logit}\{h_C(t)\} = -5 + 0.1 t,
\]

This yields a realistic increase in censoring over time.




\section{Additional Analysis Details}\label{sec:additionalAnalyses}

\subsection{Data Preprocessing and Details}\label{sec:preprocessing}

Data were obtained from the DECIFER study. We restricted the analysis to patients with treatment information and outcome information at week 52.

Baseline covariates collected in the study were comprehensive and included demographic characteristics, psychiatric history, clinical assessments, and imaging measures. Demographic variables included age, gender, socioeconomic indicators, substance use, and prior hospitalization history. Psychiatric history variables captured prior suicide-related behaviors. Clinical severity and functioning were assessed using validated instruments, including the Clinical Global Impression (CGI), Brief Psychiatric Rating Scale (BPRS), Scale for the Assessment of Negative Symptoms (SANS), Calgary Depression Rating Scale (CDRS), and quality-of-life subscales.

Variables with excessive missingness (approximately $>$50\%) were excluded a priori, including selected survey instruments (e.g., BDI, GAF, MacArthur scale). Genetic variables were also excluded due to moderate-to-high missingness (15–46\%) and limited availability across participants.

After our reduction, we find $n=52, p = 35$ covariates. An important covariate is Duration of Untreated Psychosis as noted in \cite{Goff19}. Two samples have this data missing so we remove the two yielding a final analytical set of $n=50, p =35$.
The covariates, treatment, and outcome descriptions used are summarized in \cref{tab:covariates}.

\begin{table}[ht!]
\centering
\begin{tabular}{p{4cm} p{10cm}}\toprule
\textbf{Variable} & \textbf{Description} \\
\midrule

\multicolumn{2}{l}{\textit{Treatment}} \\

Citalopram Treatment & Indicator of treatment assignment (1 = citalopram, 0 = control/placebo) \\

\midrule
\multicolumn{2}{l}{\textit{Outcome}} \\

Time to Clinically Meaningful Improvement & First time point at which the reduction in depression severity (CDRS score) from baseline exceeds a pre-specified threshold $M$ \\

\midrule
\multicolumn{2}{l}{\textit{Baseline Covariates}} \\

Age & Age at baseline assessment (years) \\

Sex & Biological sex (male or female) \\

Race & Self-reported race (White, Black, American Indian, Asian, Pacific Islander, or Other) \\

Hispanic Ethnicity & Hispanic ethnicity (yes or no) \\

Employment Status & Employment status (unemployed, full-time, or part-time) \\

Living Situation & Current residence (e.g., living alone, with family, supported housing, or structured setting) \\

Tobacco Use & Tobacco use status (none, past, or current) \\

Alcohol Use & Alcohol use status (none, past, or current) \\

Drug Use & Drug use status (none, past, or current) \\

Substance Use Disorder Diagnosis & Presence of a diagnosed substance use disorder (yes or no) \\

Number of Prior Psychiatric Hospitalizations & Total number of previous psychiatric hospitalizations \\

Suicidal Ideation at Baseline & Presence of suicidal thoughts at baseline (yes or no) \\

Legal Issues & History of legal problems (yes or no) \\

Medication Adherence (Past Month) & Self-reported adherence to medication over the past month (poor to excellent scale) \\

History of Hospitalization for Suicide Attempts & Prior hospitalization due to suicide attempts (yes or no) \\

Duration of Untreated Psychosis & Time from symptom onset to treatment initiation (weeks) \\

Baseline Quality of Life Score & Overall quality of life at baseline \\

Baseline Psychiatric Symptom Severity & Total score on the Brief Psychiatric Rating Scale (BPRS) \\

Baseline Negative Symptom Severity & Composite score from the Scale for the Assessment of Negative Symptoms (SANS) \\

Baseline Depression Severity & Total score on the Calgary Depression Rating Scale (CDRS) \\

Global Clinical Severity & Clinician-rated overall severity of illness \\

Medication Adherence Score & Self-reported adherence score (MARS scale) \\

Daily Antipsychotic Medication Use & Number of antipsychotic pills taken daily at baseline \\

\bottomrule
\end{tabular}
\caption{Summary of covariates adjusted for, treatment, and outcome. \label{tab:covariates}}
\end{table}

\subsection{Sensitivity to Time-to-Event Threshold Definitions}
We carry out sensitivity analyses to explore how different choices of thresholds $M$ affect our findings. We examine the mean and 25th percentile, finding that similar precision gains can be realized.

\begin{figure}[ht!]
\includegraphics[width=1\textwidth]{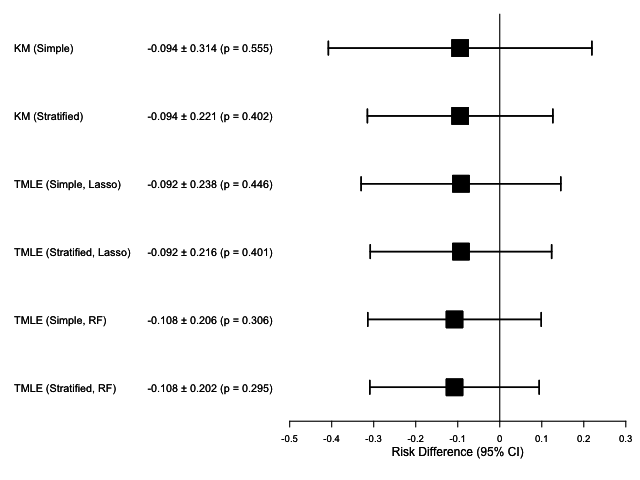}
\caption{Forest Plot of our analyses, under mean threshold. \label{fig:mean}}
\end{figure}

\begin{figure}[ht!]
\includegraphics[width=1\textwidth]{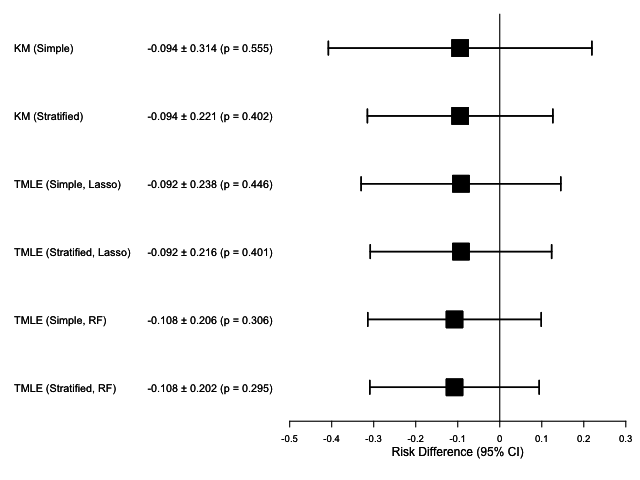}
\caption{Forest Plot of our analyses, under 25th percentile cutoff. \label{fig:t25}}
\end{figure}

\clearpage 

\section{Extensions to non Time-to-Event Outcomes}\label{sec:tmle_generalizations}
In this section, we propose methods for covariate adjustment first using Augmented Inverse Probability Weighting (AIPW), then Targeted Maximum Likelihood Estimation (TMLE). We find that our TMLE procedures offer greater precision than AIPW. While this doesn't hold in generality at a theoretical level, we present our extension in this way to demonstrate how our TMLE proposal removes a source of bias that AIPW does not account for.

Our setup is $Z_i = (X_i, A_i, Y_i(0), Y_i(1)) $ are iid from $P$. As before $W_i \subseteq X_i$, the strata variable. We observe $O_i = (X_i, A_i, Y_i(A_i))$. 

We begin with the following \textbf{Identification Assumptions.}
\begin{enumerate}[label=\bfseries (Id\arabic*)]    
    \item Consistency. $Y_i = Y(a)$ if $A_i = a$.
    \item Positivity among strata. $\prob(A_i =a \mid W_i) = p_A \geq \rho > 0, \forall i, a$.
    \item No Unmeasured Confounding, conditional on strata. 
    $O^{(n)} \perp \An \mid \Sn$. 
    \label{ass:unmeasuredConfounding}
\end{enumerate}

Our models are as follows: 
\begin{enumerate}
    \item Outcome model. $h_a(X_i)=\E[Y_i \mid X_i, A_i=a]$
    \item Propensity score model. $p_A(X_i) = \pr[A_i=1\mid X_i]$
\end{enumerate}

As before, we will allow for the propensity score to depend on covariates using a Taylor expansion to accout for this. 

\subsection{Methods}

\subsubsection{AIPW}
Our AIPW approach is as follows. We present this generalized to cross-fitting.

\begin{tcolorbox}[title=AIPW under Stratified Randomization]\label{proc:main}
\begin{enumerate}
\tightlist
        \item Split the data into $K$ folds, denoted $\vv_1 \dots \vv_K$.
        \item For $k \in [K]$,
        \begin{enumerate}
            \item Train an outcome model $\hat{h}_{a,k}$ for $a \in \{ 0, 1 \}$ using all folds but fold $k$, denoted $\vvmk$. This can include data-adaptive variable selection methods.
            \item Construct an estimate for the propensity score for strata $w \in \Sg$ denoted by
            $$ \estPik $$
            where the model is parametric and estiamted via Maximum Likelihood Estimation (MLE).
        \end{enumerate}
        \item Construct an estimator of the ATE by
            $$
            \hat{\theta}_{AIPW} = \hat{\theta}_{1,AIPW} - \hat{\theta}_{0, AIPW}$$
            for $$ \hat{\theta}_{a, AIPW} = \frac{1}{K}\sum_{k=1}^K \frac{1}{|\vvk|}\sum_{i \in \vvk} \frac{\I{A_i=a}}{\estPiki}(Y_i-\hat{h}_{a,k}(X_i)) + \hat{h}_{a,k}(X_i)$$
\end{enumerate}
\end{tcolorbox}

We make the following \textbf{AIPW Assumptions.}
\begin{enumerate}[label=\bfseries (AIPW\arabic*)]    
    \item Consistency of outcome model estimates, to some possibly misspecified outcome regression function limit $\tilde h$. 
    
    We assume that our outcome model is consistently estimated. \label{ass:consH}
    \begin{align*}
        \sqrt{\frac{1}{|\vvk|}\sum_{i \in \vvk}  (\tilde h_{a,k}(X_i) - \hat{h}_{a,k}(X_i))^2} = \opo
    \end{align*}
    \item Uniform integrability of empirical L2 norm. \label{ass:unifIntH}

    The following quantity is uniformly integrable,
    \begin{align*}
        \frac{1}{n}\sum_{i=1}^n (\tilde h_{a,k}(X_i) - \hat{h}_{a,k}(X_i))^2
    \end{align*}
\end{enumerate}

We remark that these assumptions are as \cite{VLDV24}.

We seek to show the following result:
\begin{theorem}\label{thm:aipw}
    \begin{align*}
        \frac{1}{\sqrt{|\vvk|}}\sum_{i \in \vv_k} \frac{A_i}{\estPiki} & (Y_i - \estHiok) + \estHiok \\
        &- (\frac{1}{\sqrt{|\vvk|}}\sum_{i \in \vv_k} \frac{1-A_i}{1-\estPiki} (Y_i - \estHizk) + \estHizk ) \\
        & - \trueMo - \trueMz \\
        &\qt\rightsquigarrow \mathcal{N}(0, V_{AIPW})
    \end{align*}
    where 
    $$ V_{AIPW} = V_{AIPW,1} + V_{AIPW,0} $$
    for $$V_{AIPW,1} = \frac{\V[\ytild(1)]}{p_a} + \frac{1-p_a}{p_a} \E[\V[\tilde h_{1,k}(X_i) \mid W]] + \Vmisspec(1)$$
    and $$V_{AIPW,0} = \frac{\V[\ytild(0)]}{1-p_a} + \frac{p_a}{1-p_a} \E[\V[\tilde h_{0,k}(X_i) \mid W]] + \Vmisspec(0)$$
    where $$\Vmisspec(a)=B_a \Sigma_{\beta} B_a^\top$$
    $$B_1 = \sum_w p(w)\frac{\E[ \bigg( \derivPiki \bigg)  A_i (Y_i - \trueHio) \mid W]}{p_a^2} \bigg[ \E[\frac{\partial v (X; \beta_w )}{\partial \beta}] \bigg ]^{-1} $$
    $$B_0 = \sum_w p(w) \frac{\E[ \bigg( \derivPiki \bigg)  A_i (Y_i - \trueHio) \mid W]}{p_a^2} \bigg[ \E[\frac{\partial v (X; \beta_w )}{\partial \beta}] \bigg ]^{-1} $$ 
    and $$\Sigma_{\beta}=\sum_w \V[v(X,w;\beta_w))]$$
\end{theorem}

The proof is found in \labelcref{pf:aipw} and proceeds as \cite{VLDV24}, with decomposition by strata variable.
We make a few remarks.

\begin{remark}
    \cite{Wang19} characterized the variance under stratified randomization, finding 
    $$ V = \tilde{V} - \frac{1}{p_a(1-p_a)} \E[\E[ (A-\truePi) D(Z_i) \mid W]^2]$$
    for $D$ the IF and $\tilde{V} = \E[D_{h}(Z_i)^2]$.  
    Plugging into this result for arm 1,
    $$ \tilde{V} = \V[\trueHio] + \frac{\V[\ytild(1)]}{p_a} + \frac{1-p_a}{p_a} \E[(\trueHio-\mu_1(X))^2]$$
    $$\frac{1}{p_a(1-p_a)} \E[\E[ (A-\truePi) D_h(Z_i) \mid W]^2] = \frac{1-p_a}{p_a} \E[(\trueHio-\mu_1(X))^2] $$
    Hence, our variance aligns with Theorem 1 of \cite{Wang19}, up to imbalance in $p_a$.
    
    Further, under correct model specification and $p_a=0.5$, they conclude that $\tilde{V}=V$. Under $p_a=0.5$ and correct model specification, we also find that $\tilde{V}=V$ since for arm 1
    $$ \V[\theta_{AIPW,1}]= \frac{\V[\ytild]}{p_a} + \frac{1-p_a}{p_a} \E[\V[\trueHio \mid W]] = \frac{1}{p_a} \E[\V[Y(1) \mid X]] + \V[\trueHio]$$
\end{remark}

\begin{remark}
    Estimating $p_a$ using an empirical estimate seems to offer some benefit for AIPW over using a parametric model. Namely, $\Vmisspec$ can be eliminated as $R_4$ in \cref{eq:r4} would instead be given by 
    \begin{align*}
        R_4 &= -\sqrt{|\vvk|} \sum_{w \in \Sg} \frac{1}{p_a^2}(\hat{p_a}_k(w)-p_a) \E[A_i (Y_i - \trueHio) \mid W_i=w]p(w) \\
        &= - \sum_{w \in \Sg} \frac{1}{p_a^2} \E[A_i (Y_i - \trueHio) \mid W_i=w]p(w) \frac{1}{\sqrt{|\vvk|}}\sum_{i \in \vvk} (A_i - p_a)\isS
    \end{align*}
    Under strong balancing, we know that $\sum_{i =1}^n (A_i - p_a)\isS \leq 1$. When performing data-splitting into $K$ folds, we can preserve the strata balances, meaning $R_4 = \opo$.
\end{remark}

\subsubsection{TMLE}
One may also employ TMLE to avoid the inflated variance from $\Vmisspec$ all together.
\begin{tcolorbox}[title=TMLE under Stratified Randomization]\label{proc:tmle}
\begin{enumerate}
\tightlist
        \item Split the data into $K$ folds, denoted $\vv_1 \dots \vv_K$.
        \item For $k \in [K]$,
        \begin{enumerate}
            \item Train an outcome model $\hat{h}_{a,k}$ for $a \in \{ 0, 1 \}$ using all folds but fold $k$, denoted $\vvmk$. This can include data-adaptive variable selection methods.
            \item Construct an estimate for the propensity score among each stratum variable $w \in \Sg$ as
            $$ \estPik $$
        \end{enumerate}
        \item Update initial predictions with TMLE estimator
        
        \item Construct an estimator of the ATE via
            $$ \hat{\theta}_{TMLE} = \hat{\theta}_{1,TMLE} - \hat{\theta}_{0,TMLE}$$
            for $$ \hat{\theta}_{a, TMLE} = \frac{1}{K}\sum_{k=1}^K \frac{1}{|\vvk|}\sum_{i \in \vvk} [\frac{\I{A_i=a}}{\estPiki}(Y_i-\hat{h}^{(u)}_{a,k}(X_i; \hat{\epsilon}) ) + \hat{h}^{(u)}_{a,k}(X_i; \hat{\epsilon})]$$
\end{enumerate}
\end{tcolorbox}

We will modify the assumptions and handle the expansion a bit differently than our initial estimator. We make the following \textbf{TMLE Assumptions.}
\begin{enumerate}[label=\bfseries (TMLE\arabic*),start=1]     
    \item Consistency of outcome model estimates. \label{ass:hConstTMLE}
    \begin{align*}
        \sqrt{\frac{1}{|\vvk|}\sum_{i \in \vvk}  (\tilde h_{a,k}(X_i; \tilde \epsilon)-\hat{h}^{(u)}_{a,k}(X_i; \tilde \epsilon))^2} = \opo
    \end{align*}
    \item Uniform Integrability of TMLE. \label{ass:unifIntTMLE}
    \begin{align*}
        \frac{1}{n}\sum_{i=1}^n (\htruetmlei{1}- \hbaseest{1}(X_i; \tilde \epsilon, \beta_s))^2
    \end{align*}
    \item Consistency of TMLE. \label{ass:tmleConsistency}
    \begin{align*}
        |\hat{\epsilon}-\tilde \epsilon| = \opo
    \end{align*}
\end{enumerate}

Then, we have the following result:
\begin{theorem}\label{thm:tmle}
    \begin{align*}
        \frac{1}{\sqrt{|\vvk|}}\sum_{w \in \Sg, i \in \vv_k} \frac{A_i}{\estPik} & (Y_i - \hat{h}^{(u)}_{1,k}(X_i; \hat{\epsilon})) + \hat{h}^{(u)}_{1,k}(X_i; \hat{\epsilon}) \\
        &- (\frac{1}{\sqrt{|\vvk|}}\sum_{w \in \Sg, i \in \vv_k} \frac{1-A_i}{1-\estPik} (Y_i - \hat{h}^{(u)}_{0,k}(X_i; \hat{\epsilon})) + \hat{h}^{(u)}_{0,k}(X_i; \hat{\epsilon}) ) \\
        & - \trueMo - \trueMz \\
        &\qt\rightsquigarrow \mathcal{N}(0, V_{TMLE})
    \end{align*}
    where $$ V_{TMLE} = V_{TMLE,1} + V_{TMLE,0}$$
    for
    $$ V_{TMLE,1} = \frac{\V[\ytild(1)]}{p_a} + \frac{1-p_a}{p_a} \E[\V[\tilde h_{1,k}(X_i) \mid W]]$$
    $$ V_{TMLE,0} = \frac{\V[\ytild(0)]}{1-p_a} + \frac{p_a}{1-p_a} \E[\V[\tilde h_{0,k}(X_i) \mid W]]$$
\end{theorem}

The proof is found in \labelcref{pf:tmle}.

\begin{remark}
    The variance our approach under TMLE is in general different, and not directly comparable to that of \citet{Bannick25} and \citet{Tu24}. Empirically, we find greater precision most likely because of the bias reduction from the targeting step and estimation of global quantities rather than stratum specific models.
\end{remark}

\subsection{Simulations for non-survival results}
We carry out a simulation study to verify the theoretical properties. We examine the variance, coverage, and power of the estimator, averaged over 1000 simulations.

Our simulation setup is described as follows.
\begin{align*}
    &n \in \{ 50, 100, 500, 1000 \} \\
    & w \in \{ 1, 2, 3, 4 \}, p_s = 0.25 \\
    & p_a = 0.5 \\
    & X \in \R^{d} \sim N(1, \Id_{d}) \\
    & Y = Z^\top \beta_0 + A \cdot \beta_A + \epsilon \\  
    &\frac{\V[\mu_a(X)]}{\V[\epsilon]} = SNR = 3\\
\end{align*}

We compare the following methods, 11 in total.
\begin{itemize}
    \item AIPW under simple randomization, stratified randomization, and joint calibration as in \cite{Bannick25}.
    \item ANCOVA. Correctly specified, incorrectly specified, partially misspecified, and simply using strata and treatment $(S,A)$.
    \item Difference-of-Means (DoM)
    \item TMLE under simple randomization and stratified randomization.
\end{itemize}

For visualization purposes, we do not include our ANCOVA adjustment from \cite{Tu24} on our plots. Under our simulations and analysis, we found that this approach yielded higher variances due to stratum specific nuisance estimates (as opposed to global nuisance estimates the other approaches utilize). 

The results are shown below. For variance, we see the incorrect ANCOVA models yield higher variance, followed by the correctly specified ANCOVA, DoM, AIPW methods, and finally TMLE, which yields the lowest variance. We can see reductions due to stratification for AIPW and TMLE. The AIPW calibration method of \cite{Bannick25}, while performant, results in higher variance cmopared to the other approaches.

For coverage, we see the ANCOVA models and DoM overcover by nature of their larger variances. AIPW and TMLE begin to approach the expected coverage level at $n=500$ samples and attains 95\% coverage by $n=10000$. While not shown, the cross-fitted versions attain 95\% coverage for smaller samples. Further, the DoM and the AIPW calibration method of \cite{Bannick25} result in undercoverage even at these sample sizes.

For power, incorrect ANCOVA models and DoM are severely underpowered. The AIPW and TMLE have similar power and dominate the other methods across all sample sizes, with TMLE being the most performant at smaller sample sizes. The correctly specified ANCOVA does well in comparison to incorrect ANCOVA models and DoM at moderate sample sizes (e.g. $n=500$), but performance suffers in small samples.

Interestingly, the joint calibration method from \cite{Bannick25} yield higher uncertainties and lower power compared to AIPW counterparts and TMLE. Yet, at small samples, the coverage of the AIPW calibration method is higher at lower sample sizes. 

\begin{figure}[!ht]
    \includegraphics[scale=0.8]{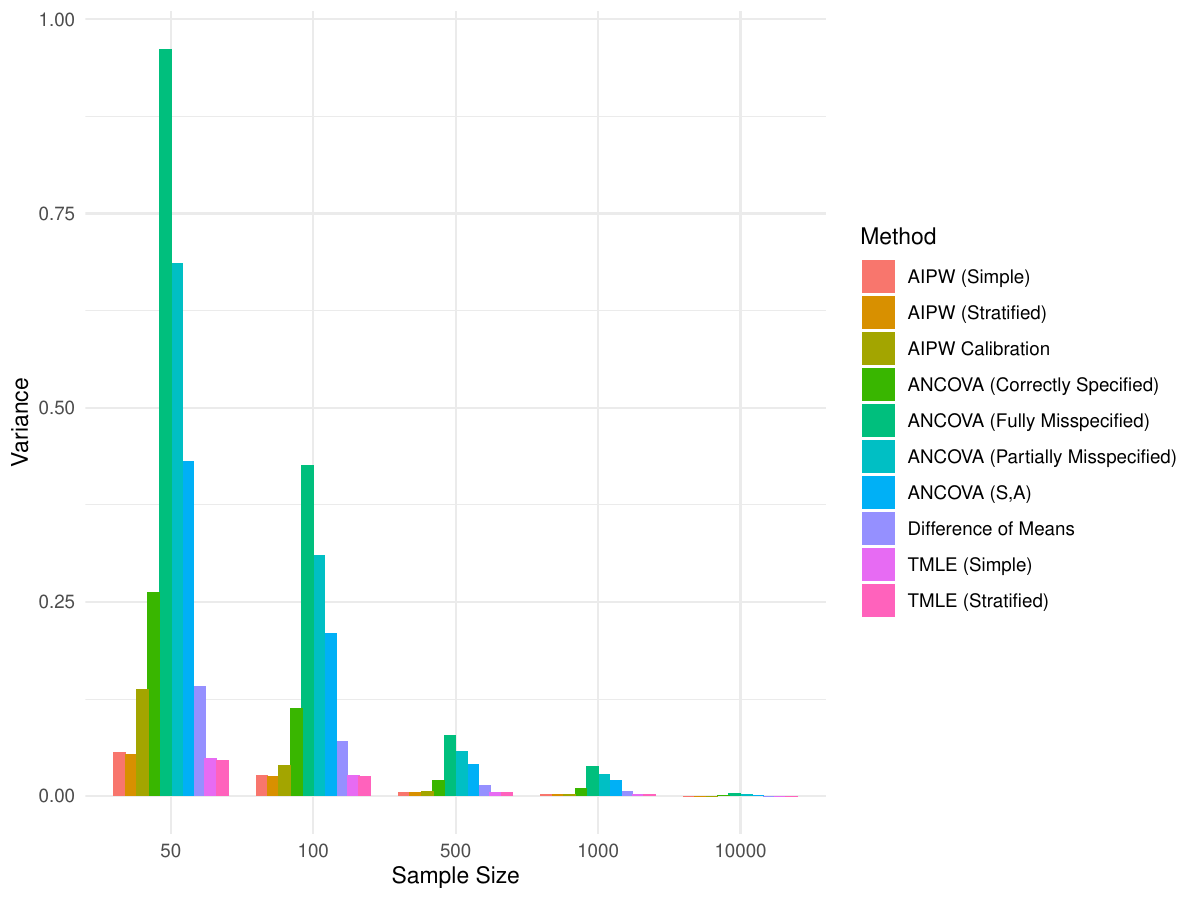}
\end{figure}

\begin{figure}[!ht]
    \includegraphics[scale=0.8]{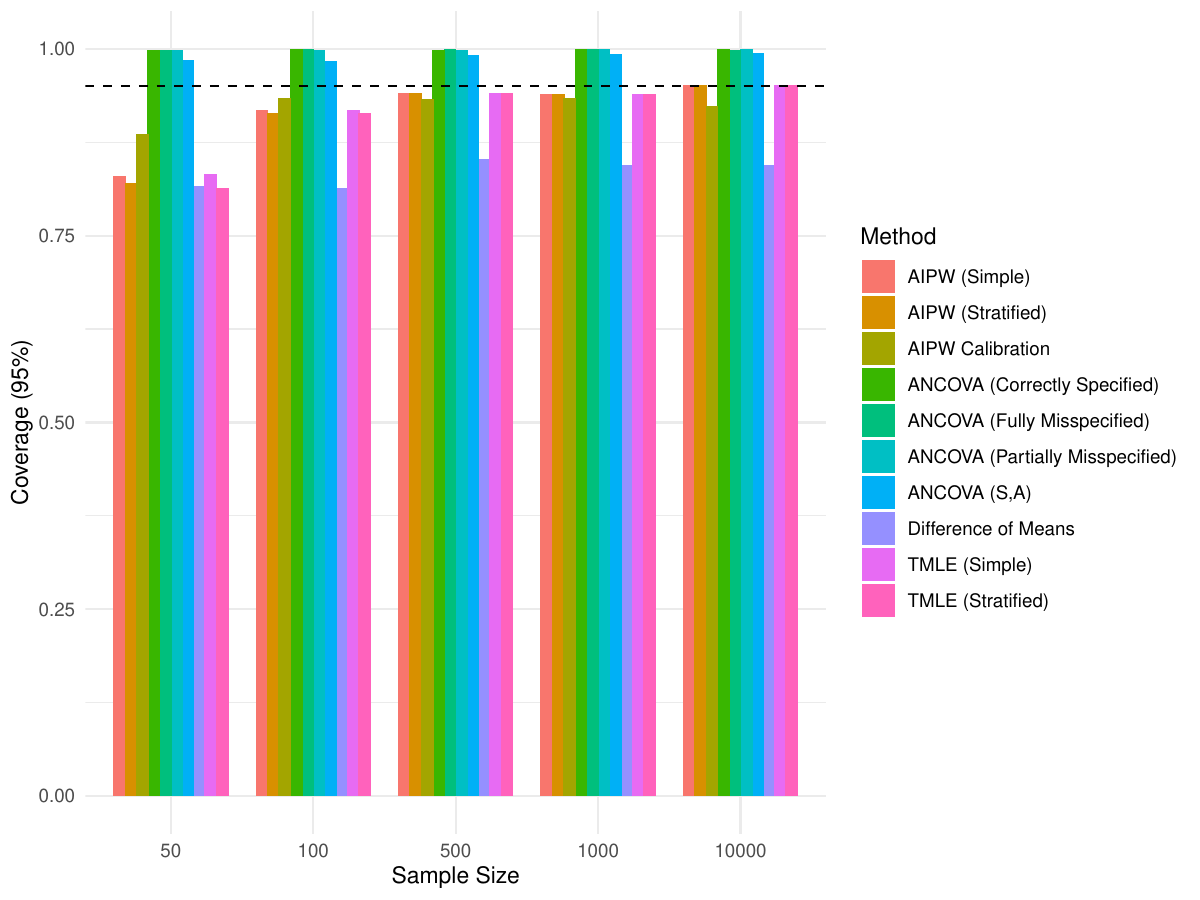}
\end{figure}

\begin{figure}[!ht]
    \includegraphics[scale=0.8]{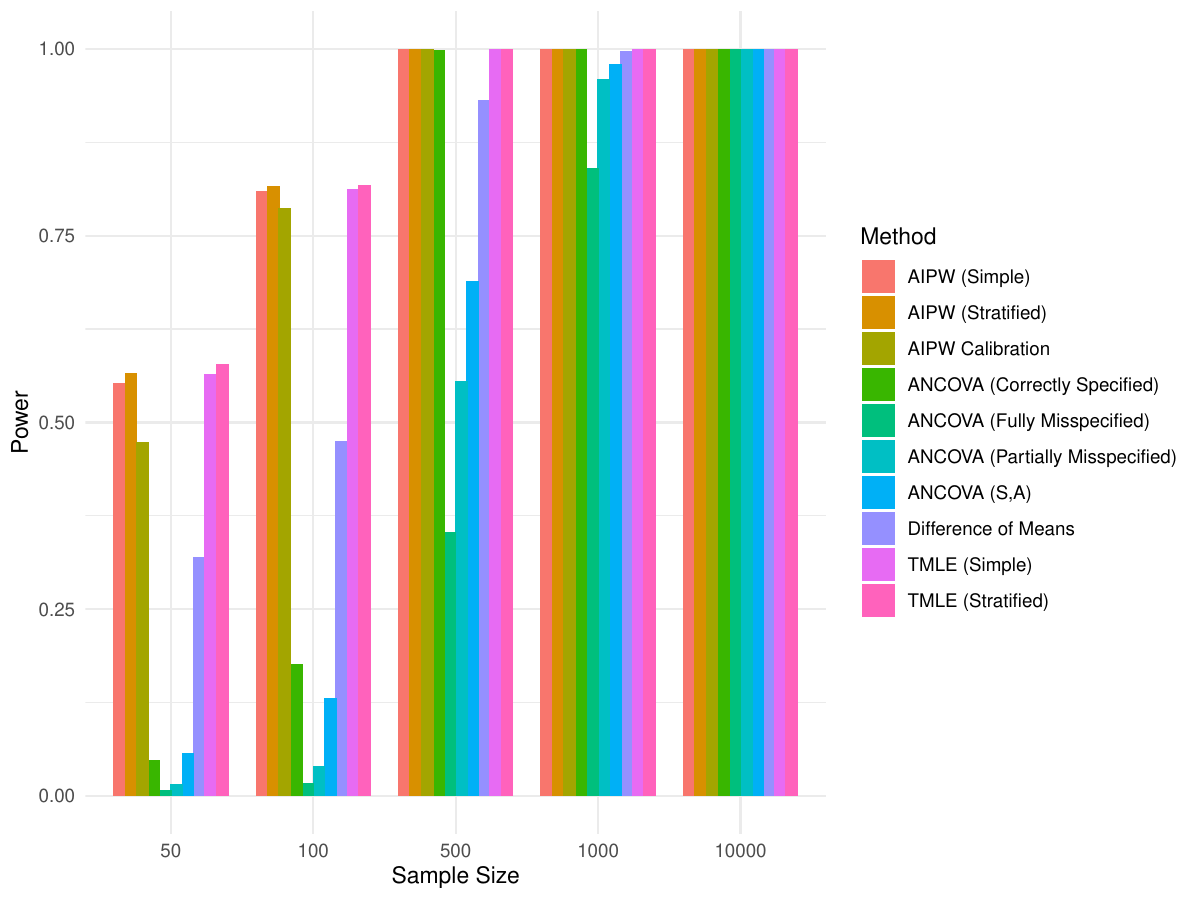}
\end{figure}

\clearpage 

\subsection{Data Analysis}
We carry out analysis on the week 52 CDRS change from baseline outcome, comparing the difference under treatment and control. 

\paragraph{Results}
Overall, we find that the estimates crosses 0 or null effect. As shown in simulation, the TMLE and our AIPW result in the highest precision gains compared to the other methods, with TMLE Stratified resulting in the lowest standard error. Below, AIPW Calibrate denote the methods of \citet{Bannick25}.

\begin{figure}[!ht]
    \includegraphics[scale=0.7]{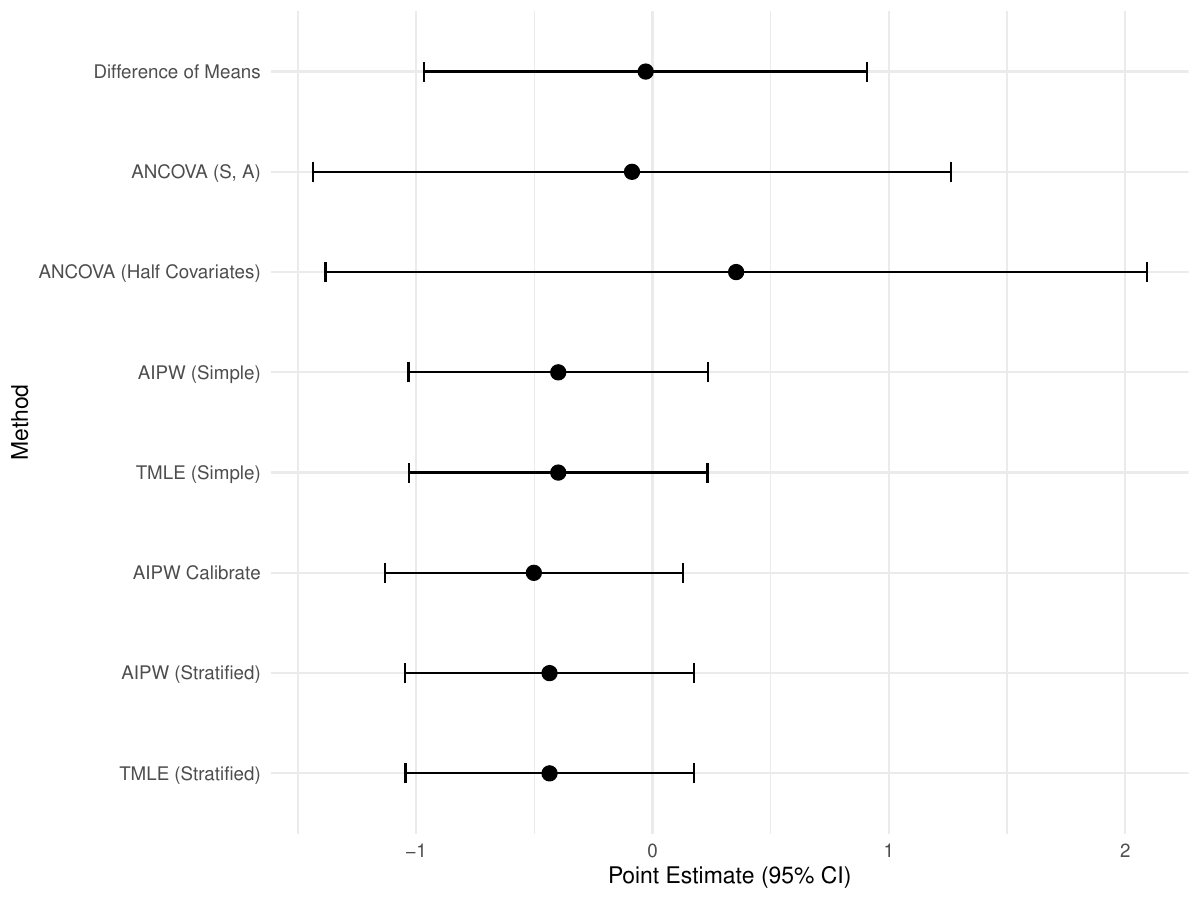}
    \label{fig:RWD_forest_plot}
\end{figure}

\begin{table}[ht!]
\centering
\begin{tabular}{||l c c c||}
\hline
Method & Estimate (95\% CI) & Standard Error & p-value \\ [0.5ex]
\hline\hline
Difference of Means
  & $-0.029\;(-0.966,\;0.908)$ & 0.478 & 0.952 \\
ANCOVA (S, A)
  & $-0.087\;(-1.438,\;1.264)$ & 0.689 & 0.900 \\
ANCOVA (Half Covariates)
  & $0.354\;(-1.384,\;2.092)$ & 0.887 & 0.692 \\
AIPW (Simple)
  & $-0.399\;(-1.032,\;0.234)$ & 0.323 & 0.223 \\
TMLE (Simple)
  & $-0.399\;(-1.030,\;0.233)$ & 0.322 & 0.222 \\
AIPW Calibration
  & $-0.502\;(-1.133,\;0.129)$ & 0.322 & 0.125 \\
AIPW (Stratified)
  & $-0.436\;(-1.047,\;0.176)$ & 0.312 & 0.169 \\
TMLE (Stratified)
  & $-0.436\;(-1.045,\;0.174)$ & 0.311 & 0.168 \\
ANCOVA (Tu)
  & $-0.301\;(-4.747,\;4.146)$ & 2.269 & 0.895 \\ [1ex]
\hline
\end{tabular}
\caption{Estimated Treatment Effects from DECIFER}
\label{fig:tab_RWD}
\end{table}

\section{Proofs for non-time-to-event outcomes}
\subsection{Proof of Theorem \texorpdfstring{\labelcref{thm:aipw}}{Theorem}}
\label{pf:aipw}

We demonstrate the result on the effect from arm 1, with results for arm 0 following similarly. Consider the decomposition from \cite{VLDV24} by $w \in \Sg$:

\begin{align*}
        \frac{1}{\sqrt{|\vvk|}}\sum_{i \in \vv_k} &[\frac{A_i}{\estPik} (Y_i - \estHiok) +  \estHiok ] \\ 
        &= \frac{1}{\sqrt{|\vvk|}}\sum_{w \in \Sg} \sum_{i \in \vv_k} \isS [\frac{A_i}{\estPik} (Y_i - \estHiok) +  \estHiok ] \\
        & = \underbrace{\frac{1}{\sqrt{|\vvk|}}\sum_{w \in \Sg, i \in \vv_k} \isS [\frac{A_i}{\truePi} (Y_i - \estHiok) + \estHiok]}_{(I)} \\
        &\quad+ \underbrace{\frac{1}{\sqrt{|\vvk|}}\sum_{w \in \Sg, i \in \vv_k} \isS [A_i (Y_i - \estHiok)(\frac{1}{\estPik}-\frac{1}{\truePi})]}_{(II)} 
\end{align*}

We now show that $II$ is asymptotically negligible and $I$ is asymptotically normal. We begin with term $II$.

\subsubsection{Term II.}
We proceed assuming sufficient regualrity conditions on the propensity score model estimation. The score will be denoted by $v$.

\begin{align*}
    \frac{1}{\sqrt{|\vvk|}} & \sum_{w \in \Sg, i \in \vvk} \isS  A_i  (Y_i - \estHiok)(\frac{1}{\estPik}-\frac{1}{\truePi})  \\
    & = \frac{1}{\sqrt{|\vvk|}} \sum_{w \in \Sg, i \in \vv_k} \isS A_i (Y_i - \trueHio)(\frac{1}{\estPik}-\frac{1}{\truePi}) \\
    &\qt + \frac{1}{\sqrt{|\vvk|}} \sum_{w \in \Sg, i \in \vv_k} \isS A_i (\trueHio - \estHiok)(\frac{1}{\estPik}-\frac{1}{\truePi})  \\
    &\overset{(a)}{\leq} - \sum_{w \in \Sg} \sqrt{|\vvk|}(\betaestk - \beta_w) \frac{1}{|\vvk|} \sum_{i \in \vvk} \isS \bigg [\frac{1}{\truePi^2} \derivPiki \bigg]  A_i (Y_i - \trueHio)  \\
    &\qt  + \sqrt{|\vvk|} \bigg(\max_{i \in \vvk, w \in \Sg} \frac{1}{\estPik \truePi} \bigg) \sum_{w \in \Sg} \sqrt{\frac{1}{|\vvk|}\sum_{i \in \vvk}  \isS (\trueHio - \estHiok)^2}  \\
    &\qt \qt \sqrt{\frac{1}{|\vvk|} \sum_{i \in \vvk} \isS (\estPik-\truePi)^2 }\\
    & \qt + \opo \\
    &\overset{(b)}{\leq} - \sum_{w \in \Sg} p(w) \frac{\E[ \bigg( \derivPiki \bigg)  A_i (Y_i - \trueHio) \mid W]}{p_a^2}  \bigg[ \E[\frac{\partial v (X; \beta_w )}{\partial \beta}] \bigg ]^{-1} [\frac{1}{\sqrt{|\vvk|}} \sum_{w \in \Sg} \sum_{i \in \vvk} v(X_i; \beta_w)] \\
    & \qt + \opo  
\end{align*}
where:
\begin{enumerate}[label=(\alph*)]
    \item follows from taking a Taylor expansion on the propensity score model for each $w \in \Sg$ on the first term and applying Cauchy-Schwarz to the second term
    \item invoking the ULLN and asymptotics from MLE (combined using Slutsky's) on the first term, and \labelcref{ass:consH} and MLE results \cite{Tsiatis} are invoked to control the second term.
\end{enumerate}

In sum, we have asymptotic contributions from 
\begin{equation}\label{eq:r4}
R_4=- \sum_{w \in \Sg} p(w) \frac{\E[ \bigg( \derivPiki \bigg)  A_i (Y_i - \trueHio) \mid W]}{p_a^2} \bigg[ \E[\frac{\partial v (X; \beta_w )}{\partial \beta}] \bigg ]^{-1} [\frac{1}{\sqrt{|\vvk|}} \sum_{w \in \Sg} \sum_{i \in \vvk} v(X_i; \beta_w)]
\end{equation}

\subsubsection{Term I.}

Consider the following expression 
\begin{align*}
    \frac{1}{\sqrt{|\vvk|}}\sum_{w \in \Sg, i \in \vv_k} \isS & \frac{A_i}{\truePi} (Y_i - \estHiok) + \estHiok - \trueMo \\
    &= \underbrace{\frac{1}{\sqrt{|\vvk|}}\sum_{w \in \Sg, i \in \vv_k} \isS\frac{A_i}{\truePi} (Y_i - \trueHio) + \trueHio - \trueMo}_{I.1}  \\
    & \qt + \underbrace{\frac{1}{\sqrt{|\vvk|}}\sum_{w \in \Sg, i \in \vv_k} \isS(\frac{A_i}{\truePi}-1) (\trueHio-\estHiok)}_{I.2}  \\
\end{align*}
We first bound $I.2$, showing it is asymptotically negligible, then demonstrate asymptotic normality using $I.1$.

\subparagraph{Term I.2}
\rck{Same way as VL24, just with strata-specific handling and CF.}

We control this by showing the bias and variance are sufficiently controlled. Because we are under stratified randomization, we should condition on $\Sn$ in addition to $\vvmk$ for cross-fitting.

\textbf{Bias:} By iterated expectations on $\vvmk, \Sn$, $\E[I.2]=0$.

\textbf{Variance:} The law of total variance yields,
\begin{align*}
    \V[I.2] & = \E[\V[\frac{1}{\sqrt{|\vvk|}}\sum_{w \in \Sg, i \in \vvk} \isS (\frac{A_i}{\truePi}-1) (\trueHio-\estHiok) \mid \vvmk, \Sn]] \\
    & \leq \E[\frac{1}{|\vvk|}\E[\sum_{w \in \Sg, i \in \vvk} \isS (\frac{A_i}{\truePi}-1)^2 (\trueHio-\estHiok)^2 \mid \vvmk, \Sn] ]\\
    &\leq \E[\frac{1}{|\vvk|} \frac{1}{\rho^2} \sum_{w \in \Sg, i \in \vvk}  \isS \E[(\trueHio-\estHiok)^2 \mid \vvmk, \Sn] ]\\
    &\lesssim \E[\frac{1}{|\vvk|} \sum_{i \in \vvk} (\trueHio-\estHiok)^2] \\
    &= \opo
\end{align*}
where the final step follows from Theorem 3.5 of \cite{Billingsley} invoking \labelcref{ass:consH} and \labelcref{ass:unifIntH}.

\subsubsection{Asymptotic Normality}


It remains to show asymptotic normality. As in \cite{Bugni18, Wang19}, we will make use of auxiliary variables conditional on the strata and construct a coupling. Let $\maSi{a}=\E[Y_i(a) \mid W_i]$ and $g_a(S_i)=\E[h^*_a(X_i) \mid W_i]$. By our work above,
\begin{align*}
    \frac{1}{\sqrt{|\vvk|}}  &\sum_{w \in \Sg, i \in \vv_k} \isS [\frac{A_i}{\estPik} (Y_i - \estHiok)  +  \estHiok] - \trueMo = I + II -\trueMo \\
    &= \frac{1}{\sqrt{|\vvk|}}\sum_{w \in \Sg, i \in \vv_k} \isS [\frac{A_i}{\truePi} (Y_i - \trueHio) + \trueHio] + II - \trueMo + \opo \\
    &= \frac{1}{\sqrt{|\vvk|}}\sum_{w \in \Sg, i \in \vv_k} \isS [\frac{A_i}{\truePi} [(Y_i - \maSi{1}) + (\maSi{1}- \gaSi{1})] \\
    &\qt + [(\gaSi{1} -\trueHio)]+\trueHio-\gaSi{1} + \gaSi{1}] -\trueMo \\
    &= \underbrace{\frac{1}{\sqrt{|\vvk|}} \sum_{w \in \Sg, i \in \vv_k} \isS [\frac{A_i}{\truePi} (Y_i - \maSi{1})]}_{R_1}  \\
    &\qt + \underbrace{\frac{1}{\sqrt{|\vvk|}}\sum_{w \in \Sg, i \in \vv_k} \isS [ \frac{A_i}{\truePi} (\maSi{1}- \gaSi{1}) + \gaSi{1}-\trueMo]}_{R_2} \\
    &\qt + \underbrace{\frac{1}{\sqrt{|\vvk|}}\sum_{w \in \Sg, i \in \vv_k} \isS  [\frac{A_i-\truePi}{\truePi}(\gaSi{1} -\trueHio)]}_{R_3} \\
    &\qt  \underbrace{- \sum_{w \in \Sg} p(w) \E[ \bigg( \derivPiki \bigg)  A_i (Y_i - \trueHio) \mid W] \bigg[ \E[\frac{\partial v (X; \beta_w )}{\partial \beta}] \bigg ]^{-1} [\frac{1}{\sqrt{|\vvk|}} \sum_{i \in \vvk} v(X_i; \beta_w)] }_{R_4} \\
    &\qt + \opo
\end{align*}

We now show the limit distribution of each term.

\paragraph{Term $R_1$}
We handle this term in a similar manner as Lemma B.1 of \cite{Bugni18}.
Let $\ytild = (Y_i(a) - \maSi{a})$ for $\maSi{a}=\E[Y_i(a) \mid W_i]$, $n_a(s)= \sum_{i=1}^n \isS \I{A_i=a}$, $F(s)=\prob(W_i < w)$, and $n(s)=\sum_i \I{W_i < w }$. Further, let $\ytilds(1), \ytilds(0)$ be i.i.d. random variables with marginal distribution equal to $(\ytild(1), \ytild(0)) \mid W_i=w$. 
We begin by reordering the sum by strata and treatment, and sum over the patients that receive treatment:
\begin{align*}
    R_{1}^* = \frac{1}{\sqrt{|\vvk|}}\sum_{w \in \Sg} \sum_{i=1+ \lfloor nF(w) \rfloor}^{\lfloor n F(w) + p_a p(w) \rfloor} \frac{1}{\truePi} \ytilds(1)
\end{align*}
It is clear that the following hold:
\begin{itemize}
    \item $R_{1}^* \rightsquigarrow \N(0, \V[\frac{\ytild(1)}{p_a}])$ by Lemma B.1 of \cite{Bugni18}, for $\sigma^2_{\ytild(1)}=\V[Y(1)-m_1(W)]$. We note that reduction of this procedure to Brownian motion under regularity conditions (see theorem 8.6.5 of \cite{Durrett}), paired with convergence of $(n(w)/n, n_1(w)/n)$ and the CMT, yields the result.
    \item $R_{1}^* \overset{d}{=} R_{1}$ since $\{ R_{1}^* \mid \Sn, \An\}  \overset{d}{=} \{ R_{1} \mid \Sn, \An \}$
\end{itemize}

\paragraph{Term $R_2$}
We consider the first part of $R_2$ given by:
\begin{align*}
    \frac{1}{\sqrt{|\vvk|}}\sum_{w \in \Sg, i \in \vv_k} & \isS \frac{A_i}{\truePi} (\mas{1}- \gas{1}) \\
    & = \sum_{w} (\mas{1}- \gas{1})[\frac{1}{\sqrt{|\vvk|}} \sum_{i \in \vvk} \frac{1}{\truePi}(A_i - \truePi)\isS  + \truePi n_1(w)] \\
    &= \sum_{w} (\mas{1}- \gas{1})[\frac{1}{\sqrt{|\vvk|}} \sum_{i \in \vvk} \frac{A_i - \truePi}{\truePi} \isS]  \\
    &\qt + \frac{1}{\sqrt{|\vvk|}}\sum_{w}  \frac{1}{\truePi}(\mas{1}- \gas{1}) n_1(w) \\ 
    &= \sum_{w} (\mas{1}- \gas{1})[\frac{1}{\sqrt{|\vvk|}} \sum_{i \in \vvk} \frac{A_i - \truePi}{\truePi} \isS ]  \\
    &\qt + \sqrt{|\vvk|} \sum_{w}  \frac{1}{\truePi} (\mas{1}- \gas{1}) \frac{n_1(w)}{|\vvk|} \\ 
\end{align*}

The second part of $R_2$ can be written as:
\begin{align*}
    \frac{1}{\sqrt{|\vvk|}}\sum_{w \in \Sg, i \in \vv_k} \gaSi{1} - \trueMo &= \sqrt{|\vvk|} \sum_w (\gas{1} \frac{n(s)}{|\vvk|}-\mas{1} p(w)) \\
\end{align*}

We then have 
\begin{align*}
    R_2 
    &= \sum_w (\mas{1} - \gas{1}) (\frac{1}{\sqrt{|\vvk|}} \sum_{i \in \vvk} \frac{A_i - \truePi}{\truePi}\isS ) + \sqrt{|\vvk|} \sum_w \mas{1}(\frac{n_1(w)}{\truePi|\vvk|} - p(w)) \\
    &\qt + \sqrt{|\vvk|} \sum_w \gas{1} (\frac{n(w)}{|\vvk|} - \frac{n_1(w)}{\truePi|\vvk|}) 
\end{align*}

We first take care of our remainder terms.
\begin{align*}
    \sqrt{|\vvk|} \sum_w \mas{1}(\frac{n_1(w)}{\truePi|\vvk|} - p(w)) &= \sqrt{|\vvk|} \sum_w \mas{1}(\frac{n_1(w)}{\truePi|\vvk|} - n(w) + n(w) - p(w))
\end{align*}

From here, it is easy to see this quantity is mean 0, and by Chebyshev's, the variance of this expression under the law of total variance decays to 0 so it is asymptotically negligible.
Similarly, we can handle the other remainder term.

The leading term, under stratified or biased coin toss, is killed off since strong balance holds: that is, $\sum_i (A_i - p_a) \isS = \Opo$ \citep{Bugni18}.

\paragraph{Term $R_3$}
\begin{align*}
    R_3 &= \frac{1}{\sqrt{|\vvk|}}\sum_{w \in \Sg, i \in \vv_k} \isS [\frac{A_i-\truePi}{\truePi}(\gaSi{1} -\trueHio)] 
\end{align*}

The summand, conditional on $\Sn$ is mean 0. The variance is given by 
\begin{align*}
    \V[\frac{A-\truePi}{\truePi} (\gas{1}-\trueHio)] = \frac{1-p_a}{p_a} \E[\V[\trueHio \mid W]]
\end{align*}


\paragraph{Term $R_4$}
Recall,
\begin{align*}
    - \sum_{w \in \Sg} p(w) \E[ \bigg( \derivPiki \bigg)  A_i (Y_i - \trueHio) \mid W] \bigg[ \E[\frac{\partial v (X; \beta_w )}{\partial \beta}] \bigg ]^{-1} [\frac{1}{\sqrt{|\vvk|}} \sum_{i \in \vvk} v(X_i; \beta_w)] 
\end{align*}
This expression is mean 0, and the variance of this unruly expression is:

$$ B\Sigma_{\beta} B^\top $$ 
for $B = \sum_w p(w) \E[ \bigg( \derivPiki \bigg)  A_i (Y_i - \trueHio) \mid W] \bigg[ \E[\frac{\partial v (X; \beta_w )}{\partial \beta}] \bigg ]^{-1} $ and $\Sigma_{\beta}=\V[v(X,w;\beta_w))]$. Note that the propensity score for some $w$ is fit conditional on $w$, so the asymptotic properties from MLE would extend here.

\subsection{Proof of Theorem \texorpdfstring{\labelcref{thm:tmle}}{Theorem}}

\label{pf:tmle}

Consider the decomposition, 
\begin{align*}
    \frac{1}{K}\sum_{k \in [K]} & \frac{1}{|\vvk|} \sum_{w \in \Sg, i \in \vvk} [\frac{A_i}{\estPik} (Y_i - \hesttmlei{1})] + \hesttmlei{1} \\
    & = \underbrace{\frac{1}{K}\sum_{k \in [K]} \frac{1}{|\vvk|} \sum_{w \in \Sg, i \in \vvk} [\frac{A_i}{\truePi} (Y_i - \hesttmlei{1})] + \hesttmlei{1} }_{I} \\
    & \qt + \underbrace{\frac{1}{K}\sum_{k \in [K]} \frac{1}{|\vvk|} \sum_{w \in \Sg, i \in \vvk} \isS [A_i (Y_i - \hesttmlei{1})(\frac{1}{\estPik}-\frac{1}{\truePi})]}_{II}  \\
\end{align*}

\subsubsection{Term I}
Then, for fold $k$, 
\begin{align*}
    \frac{1}{\sqrt{|\vvk|}} & \sum_{w \in \Sg, i \in \vvk} \isS [\frac{A_i}{\truePi} (Y_i - \hesttmlei{1}) + \hesttmlei{1}] - \trueMo \\
    &= \underbrace{\frac{1}{\sqrt{|\vvk|}} \sum_{w \in \Sg, i \in \vvk} \isS \frac{A_i}{\truePi} (Y_i - \htruetmlei{1}) + \htruetmlei{1}  - \trueMo}_{I.1} \\
    &\qt + \underbrace{\frac{1}{\sqrt{|\vvk|}} \sum_{w \in \Sg, i \in \vvk} \isS (\frac{A_i}{\truePi}-1)(\htruetmlei{1}- \hesttmlei{1})}_{I.2}
\end{align*}

Term $I.1$ is handled the same way as above noting that $\tilde h_{1,k}(X_i) = \htruetmlei{1}$. Term $I.2$ will be decomposed further. 
\begin{align*}
    \frac{1}{\sqrt{|\vvk|}} \sum_{w \in \Sg, i \in \vvk} &\isS (\frac{A_i}{\truePi}-1)(\htruetmlei{1}- \hesttmlei{1}) \\
    &= \frac{1}{\sqrt{|\vvk|}} \sum_{w \in \Sg, i \in \vvk} \isS (\frac{A_i}{\truePi}-1)(\htruetmlei{1}- \hesttmlestari{1}) \\
    &\qt + \frac{1}{\sqrt{|\vvk|}} \sum_{w \in \Sg, i \in \vvk} \isS (\frac{A_i}{\truePi}-1)(\hesttmlestari{1} - \hbaseest{1}(X_i; \hat{\epsilon}, \beta_s) \\
    & \qt + \frac{1}{\sqrt{|\vvk|}} \sum_{w \in \Sg, i \in \vvk} \isS (\frac{A_i}{\truePi}-1)(\hbaseest{1}(X_i; \hat{\epsilon}, \beta_s)- \hesttmlei{1}) \\
    &= (1) + (2) + (3)
\end{align*}

\paragraph{Term I (1)}
We will show this term is $\opo$ by Chebyshev's. 
This expression is mean 0. For the variance, note
\begin{align*}
    \V[II.1] &= \V[\frac{1}{\sqrt{|\vvk|}} \sum_{w \in \Sg, i \in \vvk} \isS (\frac{A_i}{\truePi}-1)(\htruetmlei{1}- \hesttmlestari{1})] \\
    &= \E[\V[\frac{1}{\sqrt{|\vvk|}} \sum_{w \in \Sg, i \in \vvk} \isS (\frac{A_i}{\truePi}-1)(\htruetmlei{1}- \hesttmlestari{1}) \mid \vvmk, \Sn]] \\
    &= \frac{1}{|\vvk|} \E[ \sum_{w \in \Sg} \isS \sum_{i \in \vvk} \E[(\frac{A_i}{\truePi}-1)^2 \mid \vvmk, \Sn] \\
    & \qt \cdot \E[(\htruetmlei{1}-\hbaseest{1}(X_i; \tilde \epsilon, \beta_w))^2 \mid \vvmk, \Sn] ] \\
    &\lesssim \frac{1}{|\vvk|} \E[ \sum_{w \in \Sg} \isS \sum_{i \in \vvk}  \E[(\htruetmlei{1}-\hbaseest{1}(X_i; \tilde \epsilon, \beta_w))^2 \mid \vvmk, \Sn] ] \\
    &= \sum_w p(w)\E[\frac{1}{|\vvk|}\sum_{i \in \vvk}(\htruetmlei{1}-\hbaseest{1}(X_i; \tilde \epsilon, \beta_w))^2]
\end{align*}
Given \labelcref{ass:hConstTMLE}-\labelcref{ass:unifIntTMLE}, we can show that this term is $\opo$ using Theorem 3.5 of \cite{Billingsley}. Hence, by Chebyshev's, this term is $\opo$. 

\paragraph{Term I (2)}
For this next term, we take a Taylor expansion, and bound the term using Chebyshev's again. 
A Taylor expansion yields
\begin{align*}
    \frac{1}{\sqrt{|\vvk|}} &\sum_{w \in \Sg, i \in \vvk} \isS (\frac{A_i}{\truePi}-1)(\hesttmlestari{1} - \hbaseest{1}(X_i; \hat{\epsilon}, \beta_w) \\
    &=\sqrt{|\vvk|} (\tilde \epsilon-\hat{\epsilon}) \frac{1}{|\vvk|} \sum_w \isS \sum_{i \in \vvk} (\frac{A_i}{\truePi}-1) \frac{\partial}{\partial \epsilon} \hbaseest{1}(X_i; \epsilon, \beta_w)  \mid_{\epsilon=\tilde \epsilon} + O_P(\sqrt{|\vvk|} |\tilde \epsilon-\hat{\epsilon}|^2)\\
    &= \frac{1}{|\vvk|} \E[ \sum_w \isS \E[((\frac{A_i}{\truePi}-1) \frac{\partial}{\partial \epsilon} \hbaseest{1}(X_i; \epsilon, \beta_w)  \mid_{\epsilon=\tilde \epsilon})^2 \mid \vvmk, \Sn]]
\end{align*}

\begin{align*}
    \V[\frac{1}{|\vvk|} & \sum_w \isS \sum_{i \in \vvk} (\frac{A_i}{\truePi}-1) \frac{\partial}{\partial \epsilon} \hbaseest{1}(X_i; \epsilon, \beta_w)  \mid_{\epsilon=\tilde \epsilon}] \\
    &= \E[\V[\frac{1}{|\vvk|} \sum_w \isS \sum_{i \in \vvk} (\frac{A_i}{\truePi}-1) \frac{\partial}{\partial \epsilon} \hbaseest{1}(X_i; \epsilon, \beta_w)  \mid_{\epsilon=\tilde \epsilon} \mid \vvmk, \Sn]]\\
    &= \E[\sum_w \isS \E[\frac{1}{|\vvk|^2} \sum_{i \in \vvk} \E[(\frac{A_i}{\truePi}-1)^2\mid \vvmk, \Sn] \E[\frac{\partial}{\partial \epsilon} \hbaseest{1}(X_i; \epsilon, \beta_w)  \mid_{\epsilon=\tilde \epsilon}^2 \mid \vvmk, \Sn]]] \\
    &= \sum_w p(w) \E[\frac{1}{|\vvk|^2} \sum_{i \in \vvk} \E[(\frac{A_i}{\truePi}-1)^2\mid \vvmk, \Sn] \E[\frac{\partial}{\partial \epsilon} \hbaseest{1}(X_i; \epsilon, \beta_w)  \mid_{\epsilon=\tilde \epsilon}^2 \mid \vvmk, \Sn]]
\end{align*}

The quantity is bounded, and by Chebyshev's, we have this as $O_P(\frac{1}{\sqrt{|\vvk|}})$. Combined with the other terms, we have
\begin{align*}
    II.2 &\leq \sqrt{|\vvk|} |\tilde \epsilon-\hat{\epsilon}| |\frac{1}{\sqrt{|\vvk|}} \sum_{w \in \Sg, i \in \vvk} \isS (\frac{A_i}{\truePi}-1)(\hesttmlestari{1} - \hbaseest{1}(X_i; \hat{\epsilon}, \beta_w) | \\
    & = \sqrt{|\vvk|} \opo O_P(\frac{1}{\sqrt{|\vvk|}})
\end{align*}

\paragraph{Term I (3)}
\begin{align*}
    \frac{1}{\sqrt{|\vvk|}} & \sum_{w \in \Sg, i \in \vvk} \isS (\frac{A_i}{\truePi}-1)(\hbaseest{1}(X_i; \hat{\epsilon}, \beta_w)- \hesttmlei{1}) \\
    &\qt = \sqrt{|\vvk|}(\beta_w - \betaestk) \frac{1}{|\vvk|} \sum_{w \in \Sg, i \in \vvk} [(\frac{A_i}{\truePi}-1) \frac{\partial}{\partial \beta} \hbaseest{1}(X_i; \hat{\epsilon}, \beta) \mid_{\beta=\beta_w}] \\
    & \qt \qt + \sqrt{|\vvk|} ||\beta_w-\betaestk||^2 O_P(1) \\
\end{align*}

Again, we can control the inner empirical mean $\frac{1}{|\vvk|} \sum_{w \in \Sg, i \in \vvk} [(\frac{A_i}{\truePi}-1) \frac{\partial}{\partial \beta} \hbaseest{1}(X_i; \hat{\epsilon}, \beta) \mid_{\beta=\beta_w}]$ via Chebyshev's. Additionally, we have $\sqrt{|\vvk|}(\beta_w - \betaestk)=\Opo$. By elementary M-Estimation properties, the remainder term $\sqrt{|\vvk|} \cdot ||\beta_w-\betaestk||^2 O_P(1)=\opo$.

Note that the first component can be shown to be $\opo$ by \labelcref{ass:hConstTMLE}. 
For the second component, consider as in Section A.4.1 \cite{VLDV24} taking a Taylor expansion. Then, we can simplify as follows:
\begin{align*}
    \sqrt{|\vvk|} \sum_{w \in \Sg} & (\tilde \epsilon-\hat{\epsilon}_w) \frac{1}{|\vvk|} \sum_{i \in \vvk} \isS (\frac{A_i}{\truePi}-1) \frac{d}{d\epsilon}\hat{h}_{1,k}^{(u)}(X_i ; \epsilon) \bigg|_{\epsilon=\tilde \epsilon} \\
    &\qt = \sqrt{|\vvk|} \opo O_P(\frac{1}{\sqrt{|\vvk|}}) + \opo\\
    &\qt = \opo
\end{align*} 

Hence,  $I.2 = \opo$.

\subsubsection{Term II}
Now, we return to the remainder term:
\begin{align*}
\frac{1}{|\vvk|} &\sum_{w \in \Sg, i \in \vvk}  \isS [A_i (Y_i - \hesttmlei{1})(\frac{1}{\estPik}-\frac{1}{\truePi})]\sqrt{|\vvk|} \\
& = \frac{1}{|\vvk|} \sum_{w \in \Sg, i \in \vvk} \isS [A_i (Y_i - \trueHio)(\frac{1}{\estPik}-\frac{1}{\truePi})]\sqrt{|\vvk|} \\
&+ \frac{1}{|\vvk|} \sum_{w \in \Sg, i \in \vvk} \isS [A_i (\trueHio - \hbaseest{1}(X_i; \tilde \epsilon, \beta_w))(\frac{1}{\estPik}-\frac{1}{\truePi})]\sqrt{|\vvk|} \\
&+ \frac{1}{|\vvk|} \sum_{w \in \Sg, i \in \vvk} \isS [A_i (\hbaseest{1}(X_i; \tilde \epsilon, \beta_w) - \hbaseest{1}(X_i;\hat{\epsilon}, \beta_w))(\frac{1}{\estPik}-\frac{1}{\truePi})]\sqrt{|\vvk|} \\
&+ \frac{1}{|\vvk|} \sum_{w \in \Sg, i \in \vvk} \isS [A_i (\hbaseest{1}(X_i;\hat{\epsilon}, \beta_w) - \hesttmlei{1})(\frac{1}{\estPik}-\frac{1}{\truePi})]\sqrt{|\vvk|} \\
&= (1)+(2)+(3)+(4)
\end{align*}

\paragraph{Term II (1)}
In a similar manner to \cref{eq:r4}, Term (1) can be rewritten as 
\begin{align*}
    \sum_w p(w) \E[A_i(Y_i- \trueHio) \frac{1}{\truePi^2}\bigg( \derivPiki \bigg) \mid W] \bigg[ \E[\frac{\partial v (X; \beta_w )}{\partial \beta}] \bigg ]^{-1} [\frac{1}{\sqrt{|\vvk|}} \sum_{i \in \vvk} v(X_i; \beta_w)] 
\end{align*}
which is $\opo$ as the inner expectation is in fact 0.

\paragraph{Term II (2)}
\begin{align*}
    \frac{1}{|\vvk|} & \sum_{w \in \Sg, i \in \vvk} \isS [A_i (\trueHio - \hbaseest{1}(X_i; \tilde \epsilon, \beta_w))(\frac{1}{\estPik}-\frac{1}{\truePi})]\sqrt{|\vvk|} \\
    &= \frac{1}{|\vvk|} \sum_{w \in \Sg, i \in \vvk} \isS [A_i (\trueHio - \hbaseest{1}(X_i; \tilde \epsilon, \beta_w))\\
    & \qt((\max_{i \in \vvk}\frac{1}{\truePi \estPik})\frac{\partial}{\partial \beta} p_s(X_i; \beta)|_{\beta=\beta_w}(\betaestk-\beta_w))+||\betaestk-\beta_w||^2]\sqrt{|\vvk|} \\
    & \lesssim \sqrt{|\vvk|} \sum_{w \in \Sg} \isS ||\betaestk-\beta_w|| \cdot \sqrt{\frac{1}{|\vvk|} \sum_{i \in \vvk} (\trueHio - \estHiok)^2} \\
    &= \opo
\end{align*}

\paragraph{Term II (3)}
By C.S. and MLE properties under sufficient regularity conditions,
\begin{align*}
    \frac{1}{|\vvk|} & \sum_{w \in \Sg, i \in \vvk} \isS [A_i (\hbaseest{1}(X_i; \tilde \epsilon, \beta_w) - \hbaseest{1}(X_i;\hat{\epsilon}, \beta_w))(\frac{1}{\estPik}-\frac{1}{\truePi})]\sqrt{|\vvk|} \\
    & \leq \sqrt{|\vvk|} (\max_{i \in \vvk} \frac{1}{\truePi \estPik}) \sqrt{\frac{1}{|\vvk|}\sum_{i \in \vvk} (\hbaseest{1}(X_i; \tilde \epsilon, \beta_w)-\hbaseest{1}(X_i; \hat{\epsilon}, \beta_w)^2} \\
    &\qt \sqrt{\frac{1}{|\vvk|}\sum_{i \in \vvk} (\truePi-\estPik)^2} \\
    &= \opo
\end{align*}

\paragraph{Term II (4)}
\begin{align*}
    \frac{1}{|\vvk|} &\sum_{w \in \Sg, i \in \vvk} \isS [A_i (\hbaseest{1}(X_i;\hat{\epsilon}, \beta_w) - \hesttmlei{1})(\frac{1}{\estPik}-\frac{1}{\truePi})]\sqrt{|\vvk|} \\
    & = \sum_{w \in \Sg} \sqrt{|\vvk|} (\beta_w - \betaestk) \frac{1}{|\vvk|} \sum_{i \in \vvk} [(\frac{1}{p_w(X_i; \betaestk)}-\frac{1}{\truePi}) \frac{\partial}{\partial \beta} \hbaseest{1}(X_i;\hat{\epsilon}, \beta)|_{\beta=\beta_w}] + \sqrt{|\vvk|}||\beta_w-\betaestk||^2 O_P(1) \\
\end{align*}

Since $\sqrt{|\vvk|} || \beta_w - \betaestk||^2 \Opo = \opo$, we consider the second term closer. Apply Holder's to each summand yielding
\begin{align*}
    \sum_{w \in \Sg} &\sqrt{|\vvk|} (\beta_w - \betaestk) \frac{1}{|\vvk|} \sum_{i \in \vvk} [(\frac{1}{\estPik}-\frac{1}{\truePi}) \frac{\partial}{\partial \beta} \hbaseest{1}(X_i;\hat{\epsilon}, \beta)|_{\beta=\beta_w}] \\
    & \leq \sum_{w \in \Sg}  \sqrt{|\vvk|} ||\beta_w-\betaestk||_1 ||\frac{1}{|\vvk|} \sum_{i \in \vvk} [(\frac{1}{\estPik}-\frac{1}{\truePi}) \frac{\partial}{\partial \beta} \hbaseest{1}(X_i;\hat{\epsilon}, \beta)|_{\beta=\beta_w}]||_\infty \\
    & \leq \sum_{w \in \Sg}  \sqrt{|\vvk|} ||\beta_w-\betaestk||_1 \max_{j \leq p, i \in \vvk} |\frac{\partial}{\partial \beta} \hbaseest{1}(X_i;\hat{\epsilon}, \beta)|_{\beta=\beta_w}| \cdot |\frac{1}{|\vvk|} \sum_{i \in \vvk} [(\frac{1}{\estPik}-\frac{1}{\truePi})]| \\
    & \leq \sum_{w \in \Sg}  \sqrt{|\vvk|} ||\beta_w-\betaestk||_1 \max_{j \leq p, i \in \vvk} |\frac{\partial}{\partial \beta} \hbaseest{1}(X_i;\hat{\epsilon}, \beta)|_{\beta=\beta_w}| \cdot (\max_{i \in \vvk} \frac{1}{\truePi \estPik}) \\
    & \qt \cdot |\frac{1}{|\vvk|} \sum_{i \in \vvk} |\estPik-\truePi|| \\
\end{align*}

Note by C.S.,
$|\frac{1}{|\vvk|} \sum_{i \in \vvk} |\estPik-\truePi|| \leq \sqrt{\frac{1}{|\vvk|} \sum_{i \in \vvk} [\estPik - \truePi]^2}$.

We know that $||\beta_w-\betaestk||_1 = O_P(\frac{1}{\sqrt{|\vvk|}})$. With MLE under sufficient regularity conditions, we know this expression is $\opo$.

\end{document}